\documentclass[
superscriptaddress,
amsmath, amssymb,
aps,
prl,
dblfloatfix,
nobalancelastpage,
reprint
]{revtex4-1}

\usepackage{graphicx}
\usepackage{bm}
\usepackage{url}
\usepackage{amsfonts}
\usepackage{bm}
\usepackage{caption}
\usepackage[subrefformat=parens,labelformat=parens]{subfig}
\usepackage{placeins}
\usepackage{floatrow}
\usepackage{float}

\begin{document}
	
	\preprint{APS/123-QED}
	
	\title{Photonic emulation of two-dimensional materials with antiferromagnetic order}
	
	\author{Ran Gladstein Gladstone}
	\affiliation{School of Applied and Engineering Physics, Cornell University, Ithaca, NY 14853}
	\author{Minwoo Jung}
	\affiliation{Department of Physics, Cornell University, Ithaca, NY 14853}
	\author{Yuchen Han}
	\affiliation{Department of Physics, Cornell University, Ithaca, NY 14853}
	\author{Gennady Shvets}
	\affiliation{School of Applied and Engineering Physics, Cornell University, Ithaca, NY 14853}

	\date{\today}
	
	\begin{abstract}
		We introduce an electromagnetic metamaterial -- a spin-valley photonic topological insulator (SV-PTI) -- that emulates a wide class of theoretically predicted gapped two-dimensional materials with antiferromagnetic order coupled to the valley degree of freedom. First-principles electromagnetic simulations and an analytic model reveal that two geometric parameters control the propagation properties of the chiral kink states at the domain wall between two SV-PTIs: their group velocity, polarization, and the existence/absence of topological protection. We demonstrate that a nonuniform SV-PTI structure can compress the electromagnetic energy carried by the kink states.
	\end{abstract}

	\maketitle
	
	
	The field of topological photonics~\cite{lu2014topological,shvets2017nphot} emerged almost a decade ago~\cite{raghu2008theory,wang2008reflection,wang2009observation} as an effort to emulate some of the most exotic phenomena in condensed matter physics~\cite{moore_nature10} using photons propagating in structured environments. One of the most important implications of topological physics for photonics is the existence of the robust edge states~\cite{wang2009observation,hafezi2011nphot,khanikaev2012photonic,segev_nature13,wu2015scheme,ma2017scattering,zhang_shvets_nphys17} at the domain walls separating two topologically distinct photonic structures. Using a wide variety of photonic platforms ranging from gyromagnetic crystals \cite{wang2008reflection,wang2009observation} and microring resonators \cite{hafezi2011nphot,hafezi2013imaging} to metamaterials \cite{gao2015topological} and twisted optical fibers \cite{alexeyev2007optical,ornigotti2007topological}, photonics researchers have successfully emulated most of the known topological phases of condensed matter, including the anomalous quantum Hall (AQH)~\cite{haldane_prl88}, quantum valley-Hall (QVH)~\cite{niu_prl07}, and quantum spin-Hall (QSH)~\cite{bernevig_science06} electron phases. Even more remarkably, some of the phenomena that either have not been experimentally realized, or do not have a clear analog in condensed matter physics, have been recently realized using photons. Those include topological lasers~\cite{kante_science17,segev_science18}, perfect refraction, and domain walls between heterogeneous topological phases~\cite{zhang_shvets_nphys17}.
	
	In this Letter we describe a photonic crystal (PhC) that emulates a broad class of condensed matter systems: two-dimensional (2D) topological insulators with antiferromagnetic (AFM) order coupled to a valley degree of freedom (DOF). Their topological nature is manifested by the existence of an integer spin-valley Chern number $C_{sv}=\pm 1$ \cite{macdonald_prl11,ezawa2013topological}, and of the robust kink states at the domain wall between topological insulators with opposite values of $C_{sv}$. Several 2D systems have been predicted to exhibit topological AFM order under certain conditions, such as AA-stacked bilayer graphene \cite{rakhmanov2012instabilities}, silicene with a staggered exchange term \cite{ezawa2013spin} or simultaneous doping/strain \cite{li2017strain}, single-layer graphene with an in-plane applied magnetic field \cite{aleiner2007spontaneous}, and manganese chalcogenophosphates in a monolayer form \cite{li2013coupling}. To our knowledge, none of them have been experimentally realized.
	
\begin{figure}
		\begin{center}
			
			\includegraphics[width=1\columnwidth]{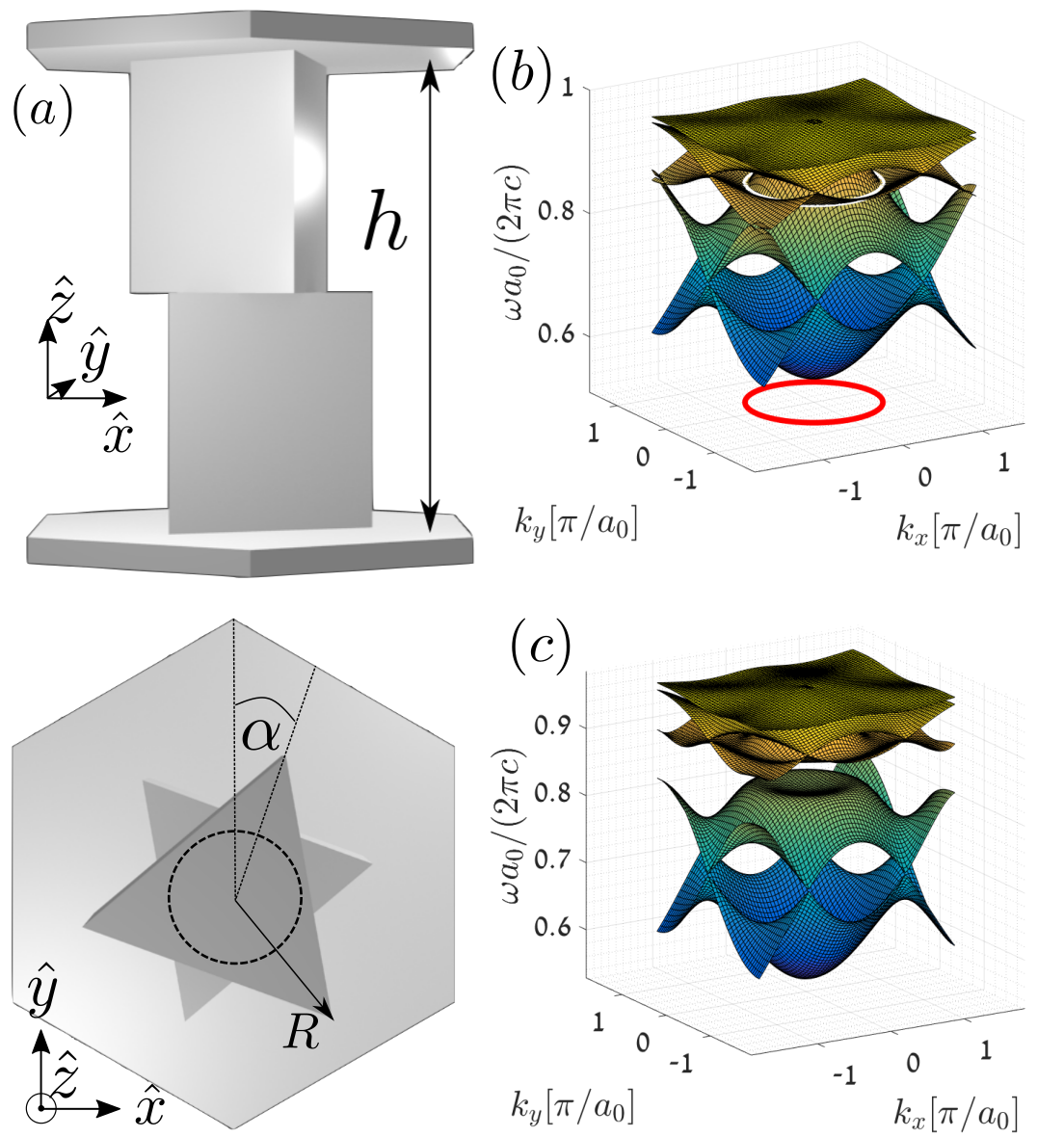}\label{fig:gammoidsGeometry}
		
		\end{center}	
		\caption{\textbf{(a)} Schematic (side and top views) and geometry definitions of the metarods used as a unit cell for an SV-PTI. Circular metarod depicted by the dashed circle represents the unperturbed photonic graphene. \textbf{(b,c)} The photonic band structures for $\alpha = 0^\circ$, no bandgap \textbf{(b)}, and $\alpha = 10^{\circ}$, complete bandgap \textbf{(c)}. White line in \textbf{(b)}: the ACNL at $\omega_{\rm bg} a_0/2\pi c \approx 0.86$. Geometric parameters: $h = 0.87a_{0}$ and $R = 0.4a_{0}$, where $a_0$ is triangular lattice period.}\label{fig:tripods}
	\end{figure}

	We show that the proposed PhC, based on a triangular array of designer (meta-) rods shaped as mutually-inverted equilateral triangular prisms (see Fig.\ref{fig:tripods}(a) for schematic and geometry definitions) made of perfect electric conductors (PEC) and sandwiched between two PEC plates, can emulate such electronic systems. By varying the two geometric parameters of the metarods  -- their size $R$ and orientation angle $\alpha$ -- one can construct a spin-valley photonic topological insulator (SV-PTI), and to capture the most salient features of the systems with valley-coupled AFM order. Those include: (i) the opening of a complete photonic bandgap opens for $\alpha \neq 0^\circ$; (ii) the existence of a spin-valley photonic topological insulator (SV-PTI) state for $R \equiv R_{\rm TP}$ characterized by an integer spin-valley Chern number $C_{sv}=\pm 1$ \cite{macdonald_prl11,ezawa2013topological}; (iii) the emergence of bandgap-crossing kink states at the domain wall between two SV-PTIs with opposite values of $C_{sv}$. The synthetic spin DOF for the $\alpha=0^\circ$ orientation is defined by the symmetry of the propagating waves with respect to $C_2$ ($180^{\circ}$) rotations around the principal axes of the PhC. In addition, we find that the bandgap-crossing kink states persist even for those geometric parameters that do not correspond to strict topological robustness, and that the latter can be removed without closing the bandgap~\cite{ezawa2013scientificreports} in the process.


	
	To understand the electromagnetic properties of the SV-PTI, we start with riangular array of circular rods (dashed line in Fig.~\ref{fig:tripods}(a), bottom) connecting the two PEC plates. Such "photonic graphene" \cite{segev_prl07} PhC (period: $a_0$, inter-plate spacing: $h$) can be designed to support a photonic band structure (PBS) consisting of doubly-degenerate Dirac cones at the $K/K^{\prime}$ points of the Brillouin zone (BZ) due to the accidental TE/TM degeneracy~\cite{ma2015guiding}. Therefore, in the $K/K^{\prime}$ valleys (i.e. for small-momentum values of $\mathbf{q} \equiv \mathbf{k} - \mathbf{K/K^{\prime}}$, where $\mathbf{K/K^{\prime}} = (\pm 4\pi /3a_0,0)$ and $\mathbf{k}\equiv(k_x,k_y)$ are the Bloch wave-vectors), wave propagation is described by a graphene-like Hamiltonian~\cite{ma2015guiding} $H_0 = \nu_{D}\left( q_{x} \hat{\tau}_{z} \hat{s}_{0}^{\prime} \hat{\sigma}_{x} + q_{y} \hat{\tau}_{0} \hat{s}_{0}^{\prime} \hat{\sigma}_{y} \right)$. Here $\hat{\sigma}_i$, $\hat{s}_i^{\prime}$, and $\hat{\tau}_i$ are the Pauli matrices associated with the orbital (LCP/RCP), synthetic $\uparrow/\downarrow$ spin (symmetric/antisymmetric mode profile, as explained below), and valley ($K/K^{\prime}$) discrete DOFs, respectively. The analog of the Dirac velocity is the group velocity $\nu_{D} \equiv  | \partial{\omega}/\partial{k}|_{\mathbf{q}=0}$.

Within the {\it low-energy} model~\cite{macdonald_prb11} (named so because its validity restricted to the valleys), the interaction Hamiltonian $H_{D}$ corresponding to the perturbation of a circular rod into two mutually-inverted triangular prisms shown in Fig.~\ref{fig:tripods}(a) can be derived from the generalized Slater perturbation theory using the unperturbed field profiles at the $K/K^{\prime}$ points~\cite{ma2017scattering}. For the simplest case of $\alpha = 0^\circ$, the metarods satisfying three symmetries: $C'_2$ out-of-plane flip ($x\rightarrow x$, $y \rightarrow -y$, $z \rightarrow -z$), $\mathcal{P}$ full-inversion ($x\rightarrow -x$, $y \rightarrow -y$, $z \rightarrow -z$), and $C_3$ rotation. The perturbed Hamiltonian is given by $H_{D} = \Delta_D \hat{\tau}_{0} \hat{s}_{z}^{\prime} \hat{\sigma}_{0}$ (see the SOM for the full derivation and calculation of the coupling constant $\Delta_D(R)$). The $s_z^{\prime}$ Pauli matrix is diagonal in the basis of the $\uparrow/\downarrow$ spin states defined as symmetric (S) and antisymmetric (A) modes with respect to the $C'_2$ flip symmetry of the structure. No bandgap exists for $\alpha=0$, and the S/A bands cross in the $\omega = {\rm const} = \omega_{\rm bg}$ plane forming an accidental-crossing nodal line (ACNL), which can be single- or multiple-connected. An example of a single-connected ACNL ($h=0.87 a_0$, $R=0.4a_0$, and $\omega_{\rm bg} a_0/2\pi c \approx 0.86$) is shown in Fig.~\ref{fig:tripods}(b) as a white line. In general, the shape of the ACNL, calculated as the solutions of $\delta \omega(\mathbf{k}) = 0$ (where $\delta \omega (\mathbf{k})$ is the frequency difference between the S/A bands), is determined by $R$. For example, the ACNL collapses into the $K/K^{\prime}$ points for $R=R_{\rm TP}$, where $R_{\rm TP} \approx 0.26 a_0$ for $h=0.87 a_0$.
	
For $\alpha \neq 0$, a complete bandgap emerges around the ACNL as shown in Fig.~\ref{fig:tripods}(c). Because the perturbation preserves the $\mathcal{P}$ and $C_3$ symmetries for arbitrary $\alpha$, it can be shown (see the SOM for detailed derivation) that it does not gap the Dirac cones at the $K/K^{\prime}$ points, yet gaps the ACNL. The mathematical form of the Hamiltonian $H_{\alpha}$ corresponding metarods' rotation by $\alpha$ can be anticipated by requiring that $H_{\alpha}$ couples the upper band of the antisymmetric Dirac cone to the lower band of the symmetric Dirac cone: $H_{\alpha}^{(K)} = \Delta_{\alpha} \hat{s}_{x}^{\prime} \hat{\sigma}_{z}$. The interaction Hamiltonian is expanded to encompass the valley DOF by demanding that it is invariant under inversion, to obtain $H_{\alpha} = \Delta_{\alpha} \hat{\tau}_{0} \hat{s}_{x}^{\prime} \hat{\sigma}_{z}$. For notational convenience, we change the spin-state basis using a unitary rotation operation $\hat{s} = \exp{(i\pi \hat{s}_y^{\prime} /4)} \hat{s}^{\prime} \exp{(-i\pi \hat{s}_y^{\prime} /4)}$ (see the SOM for details), and arrive at the expression for the low-energy interaction Hamiltonian $H=H_0 + H_D + H_{\alpha}$:
	\begin{equation}\label{eq:valley_hamiltonian}
	H(\mathbf{q}) = H_0(\mathbf{q}) + \Delta_D\hat{\tau}_{0} \hat{s}_{x}\hat{\sigma}_{0} - \Delta_{\alpha} \hat{\tau}_{0} \hat{s}_{z} \hat{\sigma}_{z},
	\end{equation}
where $\Delta_{\alpha} \equiv \Delta_{\alpha}(\alpha,R)$ is a monotonic function of $-30^{\circ}<\alpha <30^{\circ}$ satisfying $\Delta_{\alpha}(0,R)=0$.
	
In the special case of $\Delta_D = 0$, the Hamiltonian $H$ is readily recognizable as describing an AFM-ordered graphene with spin-dependent mass~\cite{li2013coupling,ezawa2013spin,ezawa2013topological}, which is known to be an SV topological insulator possessing a spin-valley index given by $u_{sv} = s \cdot \tau$. Here $\tau = \pm 1$ corresponds to the $K(K^{\prime})$ valleys, and $s = \pm 1$ corresponds to the $\uparrow (\downarrow)$ spin states. This analogy motivates the designation of our photonic structure as an SV-PTI. Because the below-bandgap propagation bands of the SV-PTI possess topological invariants $2C_{s_z}^{\tau} = u_{sv} \textup{sign} \left(\Delta_{\alpha}\right)$ that can be combined into spin-valley Chern numbers $C_{sv} = \pm 1$\cite{ezawa2013topological,li2013coupling}, the edge-bulk correspondence principle~\cite{ezawa2013topological} can be used to determine the number of spin-valley polarized kink states~\cite{macdonald_prb11,zhang_shvets_nphys17} at the domain wall separating the SV phases with the opposite signs of $C_{sv}$.
	
\begin{figure}
	\begin{center}
		\includegraphics[width=0.9\linewidth]{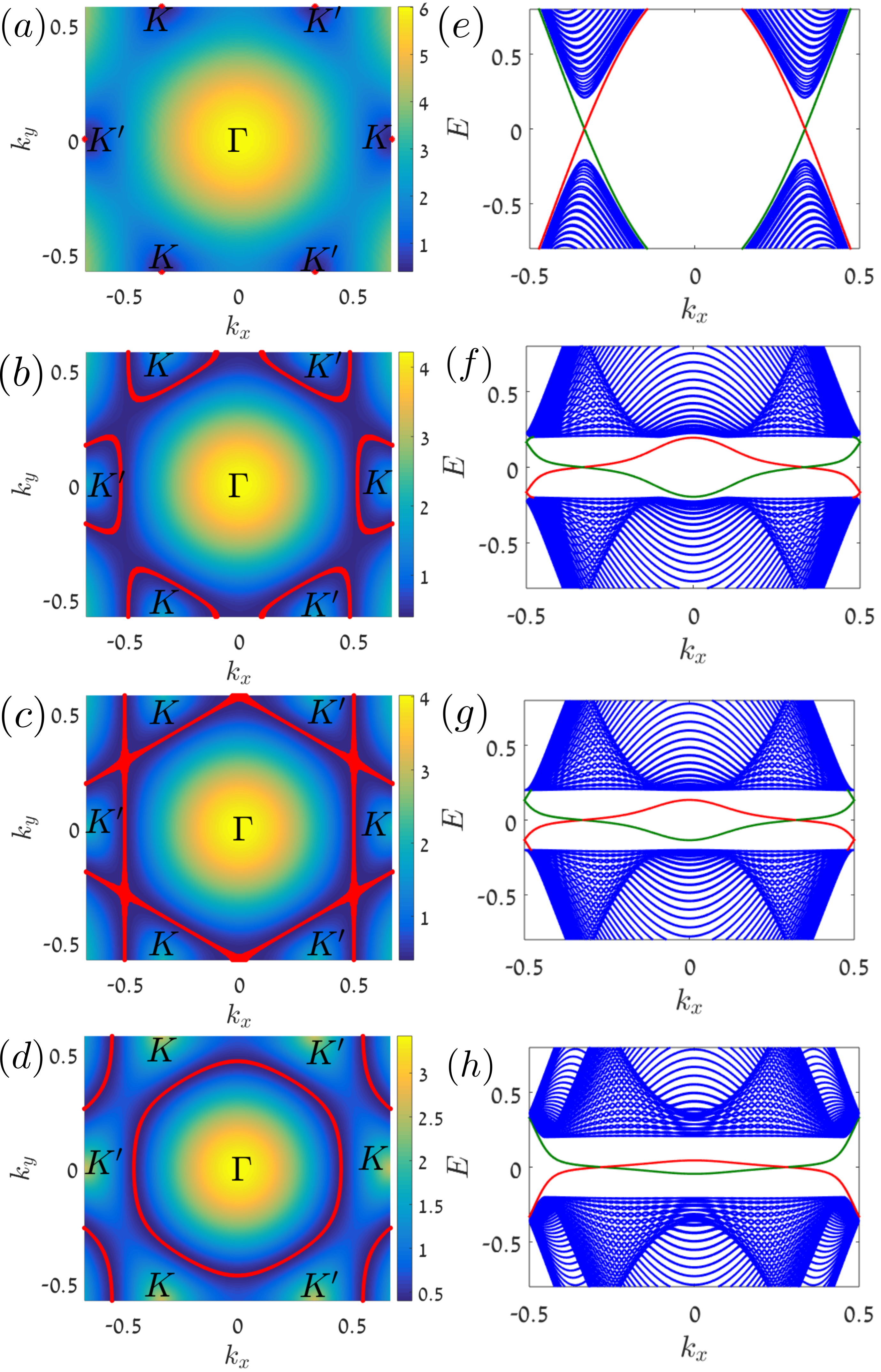}\label{fig:bulkDisp_0_3}
	\end{center}	
	\caption{Properties of the bulk and kink states in the tight-binding model with $t=1$, $\Delta_{\alpha}=0.2$, and variable $\Delta_D$. \textbf{(a)-(d)} The frequency difference $\delta\omega({\mathbf{k}})$ for the propagation bands for $\Delta_D=0.3$, $0.9$, $1.0$, and $1.3$, respectively. Red lines: minimum band separation curves defined by $\nabla_{\mathbf{k}} \delta\omega = 0$. \textbf{(e)-(h)} Bulk and kink states for two adjacent domains with opposite signs of $\Delta_{\alpha}$ separated by a zigzag-shaped domain wall. The green and red colors: mutually orthogonal (non-interacting) kink modes. \label{fig:analyticBandStructure}}
\end{figure}

	However, the effective Hamiltonian describing the SV-PTI contains an extra term proportional to $\Delta_D$, which removes the degeneracy of the Dirac cones corresponding to $s_z = \pm 1$. The spin-valley DOF is no longer conserved for $\Delta_D \neq 0$ because the Hamiltonian given by Eq.~(\ref{eq:valley_hamiltonian}) is no longer separable into two independent spin blocks. Therefore, the spin-valley Chern number $C_{sv}$ loses its quantization~\cite{ezawa2013scientificreports}, and the existence of the kink states is unknown for this general case. Furthermore, the low-energy model loses validity when finite $\Delta_D$ pushes the ACNL away from the valleys to encircling the $\Gamma$ point (see Fig.~\ref{fig:tripods}(b)). Therefore, we extend the low-energy model to the entire BZ by introducing a tight-binding model that possesses the same point group symmetry and has the same behavior in the vicinity of the $K/K'$ points as described by Eq.~(\ref{eq:valley_hamiltonian}). The tight-binding model, recently found~\cite{rakhmanov2012instabilities} (See SOM for the mapping between different notations) to describe the AA-stacked bilayer graphene with AFM order, is represented by the following effective momentum-space Hamiltonian matrix:
	\begin{equation}
	\label{eq:tb_model}
	H\left(\mathbf{k} \right) = \begin{pmatrix}
	\Delta_\alpha & -tf_{\mathbf{k}} & -\Delta_D &0 \\
	-tf^{*}_\mathbf{k} & -\Delta_\alpha & 0 & -\Delta_D\\
	-\Delta_D & 0 & -\Delta_\alpha& -tf_{\mathbf{k}}\\
	0 & -\Delta_D & -tf^{*}_{\mathbf{k}} &\Delta_\alpha
	\end{pmatrix},
	\end{equation}
	where the spin-orbit basis for the above matrix is
	\begin{equation}
	\psi_{\mathbf{k}} = \begin{pmatrix}
	\textrm{RCP, } \uparrow \\
	\textrm{LCP, } \uparrow\\
	\textrm{RCP, } \downarrow\\
	\textrm{LCP, } \downarrow
	\end{pmatrix}, \enspace f_{\mathbf{k}} = 1+2\cos\left(k_x a_0 \right)\exp\left( i\sqrt{3}{k_y a_0}\right),
	\end{equation}
where $t \equiv \nu_{D}/\sqrt{3}$ are real numbers, and $\uparrow$ ($\downarrow$) label the spin-up (spin-down) modes, respectively. To make the connection with the SV-PTI, we calculate the PBS of the tight-binding model and plot the color-coded $\delta \omega({\mathbf{k}})$ in Figs.~\ref{fig:analyticBandStructure}(a)-(d)  for several values of $\Delta_D$. The minimal band separation (MBS) curves, defined as the local minima of $\delta \omega({\mathbf{k}})$, are plotted as red lines. The geometry of these curves dramatically changes with $\Delta_D$: from encircling the valleys (for $\Delta_D < t$) to encircling the $\Gamma$-point (for $\Delta_D > t$).
	
Next, we investigate the relation between the geometry of the MBS curves and the existence of bandgap-crossing kink states.  propagating along the zigzag-shaped domain wall (chosen to be parallel to the $x-$axis) between two bulk with opposite signs of $\Delta_{\alpha}$. The existence of such states is one of the key properties of the bulk phases described by Eq.(\ref{eq:tb_model}). The dispersion relations $\omega(k_x)$ for the kink states are plotted in Figs.~\ref{fig:analyticBandStructure}(e-h) for the corresponding values of $\Delta_D$. Even though the kink states exist for most values of $\Delta_D$, they undergo a dramatic change: from topologically-robust (Fig.~\ref{fig:analyticBandStructure}(e) for $\Delta_D = 0$) to topologically-trivial bandgap-crossing (Fig.~\ref{fig:analyticBandStructure}(f-g) for $0 < \Delta_D \leq t$) to bulk-attached (Fig.~\ref{fig:analyticBandStructure}(h)for $0\Delta_D > t$). The transition from bandgap-crossing to bulk-attached kink modes happens precisely for the same value of $\Delta_D = t$ as the transition from the multiple MBS curves encircling the $K/K^{\prime}$ points to a single MBS curve encircling the $\Gamma$ point of the BZ. Note that the existence of the band-crossing kink states is insufficient for making them topologically robust. The latter condition is violated for finite $\Delta_D$ because the $\uparrow / \downarrow$ spin states are no longer conserved.



\begin{figure}
		\begin{center}
			\includegraphics[width=1\columnwidth]{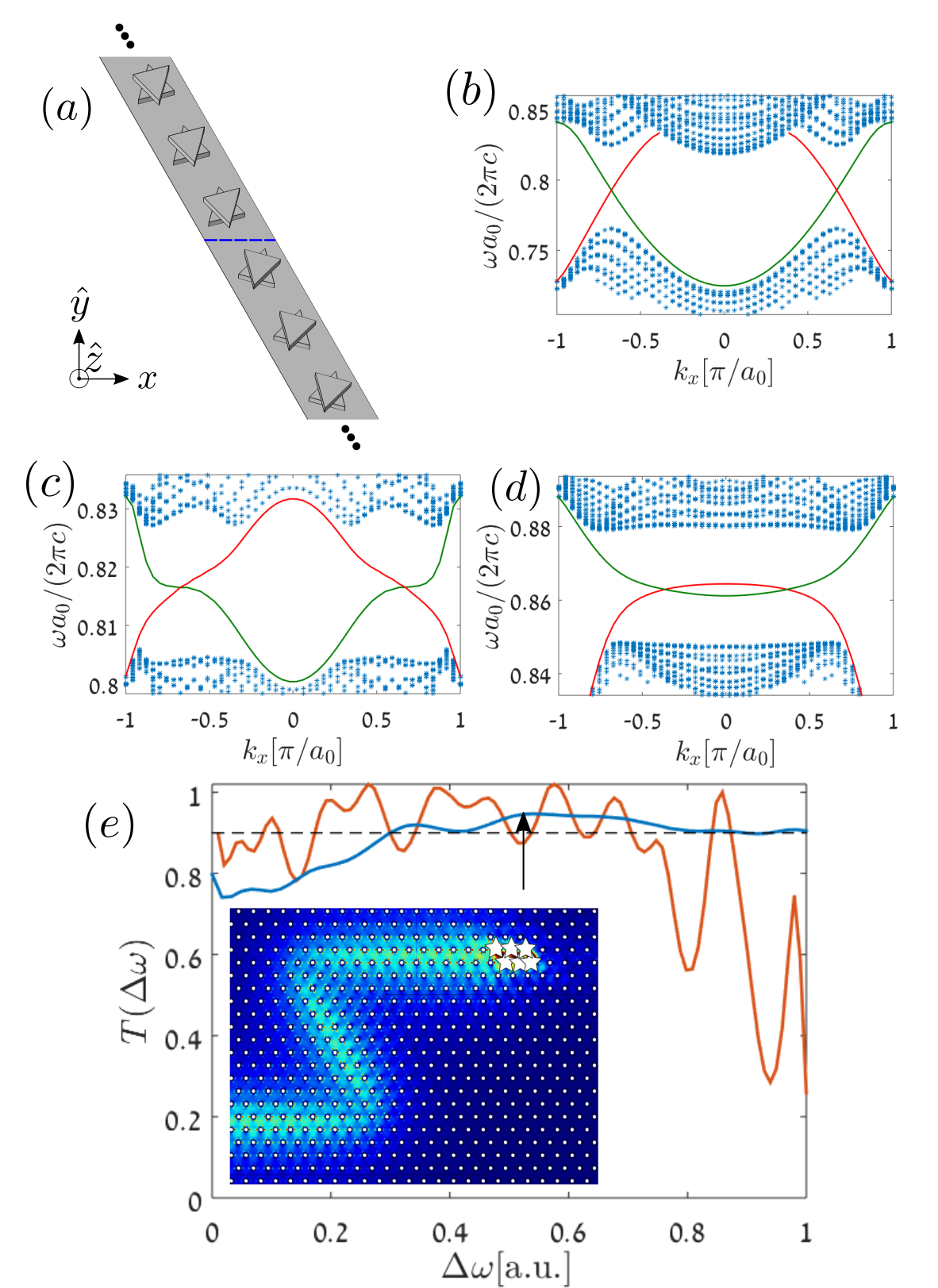}
			
		\end{center}
		
		\caption{\textbf{(a)} Schematic of the domain wall (blue line) formed by interfacing two domains with opposite signs of $\alpha$. \textbf{(b-d)} 1D band dispersions along the domain wall calculated for \textbf{(b)} $R=0.261a_0 \equiv R_{\rm TP}$ $/$ $\alpha = 30^{\circ}$ , \textbf{(c)} $R=1.26 R_{TP}$ $/$ $\alpha = 5^\circ$, and \textbf{(d)} $R= 1.53 R_{TP}$ $/$ $\alpha = 5^\circ$ ($h = 0.87a_0$ for all three cases); the bulk modes are displayed in blue dots, and the kink modes are in red(symmetric) or green(anti-symmetric) lines. \textbf{(e)} Transmission spectra $T(\omega)$ along the Z-shaped sharp turns for the topogically protected case shown in Fig. 3(b) (blue) and for a slightly perturbed structure $R=1.07 R_{TP}$ (orange). The frequency is scaled so that $\Delta \omega=0$ and $1$ corrresponds to the band gap edges, and the black dashed line ($T=0.9$) serves as a guide to the eye. Inset: A kink mode at $\omega a_0/(2\pi c) \approx 0.79$ (black arrow) is launched by phased dipoles, and propagates along the Z-shaped sharp turns with negligible reflection (colormap: $|\mathbf{E}|$).}
		\label{fig:supercellSimulations}
	\end{figure}

	These predictions of the tight-binding model suggest that similar transitions between band-crossing and bulk-attached kink states may also occur at the domain wall between two SV-PTI domains shown in Fig.~\ref{fig:supercellSimulations}(a). The two opposite signs of $\Delta_{\alpha}$ in the two domains are achieved by using double-triangle metarods with opposite signs of $\alpha$ as shown in Fig.~\ref{fig:supercellSimulations}(a) and Fig.S3(b). The first-principles (COMSOL) electromagnetic simulations of thus constructed supercell containing a domain wall reveal bandgap-crossing kink modes with linear dispersion for $R=R_{\rm TP}$ and $\alpha = 30^{\circ}$, i.e. corresponding to $\Delta_D = 0$. However, for other sizes/orientations of the metarods (e.g., $(R = 1.26 R_{TP},\alpha = 5^\circ)$ and $(R=1.53 R_{TP},\alpha = 5^\circ)$ in Figs.~\ref{fig:supercellSimulations}(c,d), respectively), the kink modes lose their linear dispersion and their topological robustness. Moreover, the kink states no longer cross the entire bandgap for $R=1.53 R_{TP}$ as shown in Figs.~\ref{fig:supercellSimulations}(d).  Qualitative similarities between the band structures of a realistic SV-PTIs shown in Figs.~\ref{fig:supercellSimulations}(b-d), and those of the tight-binding model shown in Figs.2(e-h), are apparent.
	
	Topological robustness of the kink states was numerically investigated by launching them along the path containing two sharp bends as shown in Fig.~\ref{fig:supercellSimulations}(e), and calculating the ratio of time averaged energy density spectrum after and before the two sharp bends, which we define as the transmission $T(\omega)$. For the metarod parameters corresponding to Fig.~\ref{fig:supercellSimulations}(b), i.e. for the topologically protected kink states, essentially no backscattering is observed across the entire bandgap (spectrally-flat $T(\omega)>0.9$ for $66\%$ of the bandgap frequencies, where $\Delta \omega = 0$ corresponds to the normalized frequency of $0.77$ and $\Delta \omega = 1$ to $0.81$), confirming topological protection due to spin-valley conservation. An interesting example of $T(\omega)$, shown in Fig.~\ref{fig:supercellSimulations}(e) for $R=0.28a_0$, occurs when topological protection is lost. We find $T(\omega)$ to become resonant, while still attaining high values for most of the band gap frequencies ($\Delta \omega = 0$ corresponds to the normalized frequency of $0.77$ and $\Delta \omega = 1$ to $0.83$). For larger values of $R$ the resonance effects become even more prominent, causing Fabri-Perot behavior of the energy density ratio, due to the presence of the two scattering $120^\circ$ bends.

	\begin{figure}
		\centering
		
		\includegraphics[width=0.8\columnwidth]{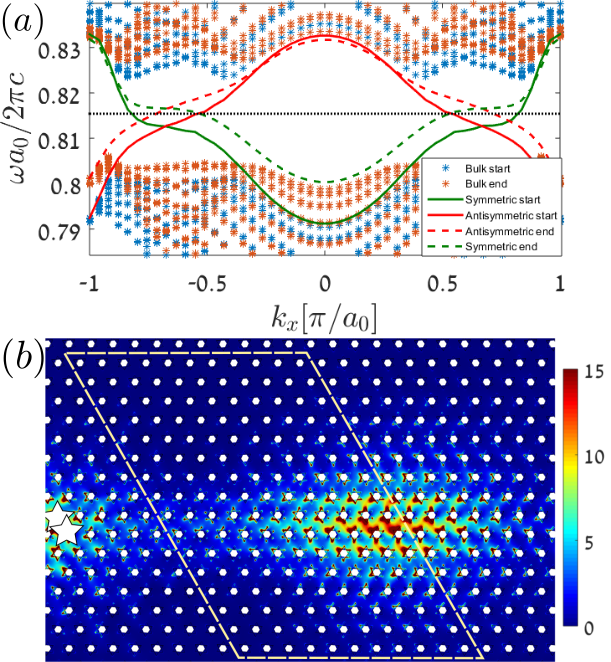}\label{fig:defectTP}
		\caption{\textbf{(a)} 1D band dispersions for the start structure ($R = 1.23 R_{TP}$; blue dots and solid lines) and the end structure ($R = 1.26 R_{TP}$; orange dots and dashed lines) of the white dashed region shown in \textbf{(b)}; the black dotted line ($\omega^{(\rm s)} a_0/(2\pi c) \approx 0.816$) denotes the operating frequency. The metarods parameters of both cases are the same as in Fig. \ref{fig:supercellSimulations}(c) except $R$. \textbf{(b)} Adiabatic change of the group velocity of kink states (colormap: energy density); the white dashed lines mark the region of adiabatic change in the triangle size $1.23 R_{TP} < R < 1.26 R_{TP}$.}
		\label{fig:drivenSim}
	\end{figure}
	
	
	Below we demonstrate how the dispersion engineering of the SV-PTI kink states demonstrated above can be used to control their group velocities $v_g \equiv \partial \omega/\partial k_x$. The resulting compression of electromagnetic energy can be potentially used for a variety of applications, such as high power microwave amplifiers \cite{8240944}. Specifically, we observe that, for $R=R_{\rm crit} \approx 0.33a_0$, the dispersion relation of symmetric kink state becomes nearly flat at $k_x \approx k_{x,0}$ (near the center of the bandgap, at $\omega^{(\rm s)} a_0/(2\pi c) \approx 0.816$) as shown in Fig.~\ref{fig:supercellSimulations}(c). For the above metarod parameters, the lowest group velocity of $v_{g1}^{({\rm s})} \approx 5\times 10^{-3} c$ is obtained. This is almost two orders of magnitude smaller than the group velocities $v_g^{({\rm SH})} \sim 0.4c$ that were demonstrated for SH-PTIs \cite{ma2015guiding,lai2016scientificreports}, and for the frequencies close to the upper propagation band at $\omega^{(\rm f)} a_0/(2\pi c) \approx 0.82$.   Note that, because of the asymmetric influence of the higher propagation bands effect, this dramatic slowing down is achieved for mostly one of the two kink states.
	
	
	Therefore, by slowly changing geometric dimensions of the metarods along the propagation path of the kink modes, it should be possible to adiabatically transfer the kink modes from the fast-propagating to slow-propagating regime. Specifically, we have used the triangles of size $R$ as the tuning parameter to demonstrate energy compression at $\omega = \omega^{(\rm s)}$. For that frequency, the group velocity is expected to change from $v_{g1}^{({\rm f})} \approx 20 v_{g1}^{({\rm s})}$ to $v_{g1}^{({\rm s})}$ as shown in Fig.~\ref{fig:drivenSim}(a), resulting in a proportional energy density compression. The energy density compression factor shown in Fig.~\ref{fig:drivenSim}(b) is $\approx 6$, i.e. somewhat smaller than one would expect from the group velocity considerations. There are several effects that are potentially responsible for it: the finite-length effects (the dispersion relations strictly hold only for infinite systems), the small but finite excitation of the "fast" antisymmetric kink state, and the loss of numerical accuracy for very small group velocities. However, Fig.~\ref{fig:drivenSim}(b) provides a clear qualitative demonstration of energy density manipulation using edge states that originate from a PTI. The adiabatic change of the metarods does not cause an increase in reflection compared to a similar platform with constant metarods dimensions and is estimated at less than $5\%$ of the energy coupling into the system.
	
	In conclusion we have demonstrated using a first principles simulation and an analytic model a photonic crystal that exhibits a full range of kink state dispersion curves that range from topologically protected   spin-valley Hall phase kink states to trivial kink states. Surprisingly, the kink states survive this transition to a trivial insulator despite the breaking of the valley degree of freedom and spin degeneracy but topological protection is very quickly lost. Furthermore, this photonic crystal emulates the Hamiltonian of topological AFM electronic systems, enabling future research into their interesting properties as well as allowing to test the validity of the valley degree of freedom for topological insulators. We demonstrated a slow light application of our system and discussed its potential for other microwave applications. Our system is experimentally viable in the microwave part of the spectrum and has a clear transmission signature that can be verified.
	
	This work was supported by the Army Research Office
	(ARO) under Grant No. W911NF-16-1-0319, and by the
	National Science Foundation (NSF) under Grants No. DMR1741788 and No. DMR-1719875. M. J. was also supported
	in part by Cornell Fellowship and in part by the Kwanjeong
	Fellowship from the Kwanjeong Educational Foundation.
	
	\FloatBarrier

	\pagebreak
	\widetext
	\begin{center}
		\textbf{\large Supplemental Material: Photonic emulation of two-dimensional materials with antiferromagnetic order}
	\end{center}
	\setcounter{equation}{0}
	\setcounter{figure}{0}
	\setcounter{table}{0}
	\makeatletter
	\renewcommand{\theequation}{S\arabic{equation}}
	\renewcommand{\thefigure}{S\arabic{figure}}
	
		\section{Definition of the symmetric and antisymmetric modes}
\begin{figure}[b]
		\centering
		\sidesubfloat[]{\includegraphics[width=0.2\columnwidth]{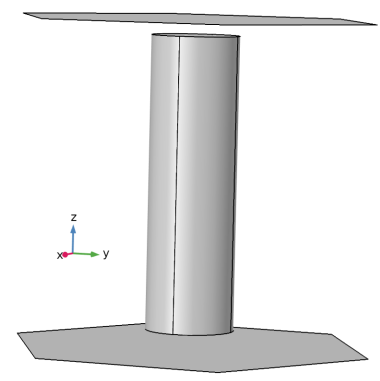}\label{fig:unperturbedPhC}}
		\sidesubfloat[]{\includegraphics[width=0.2\columnwidth]{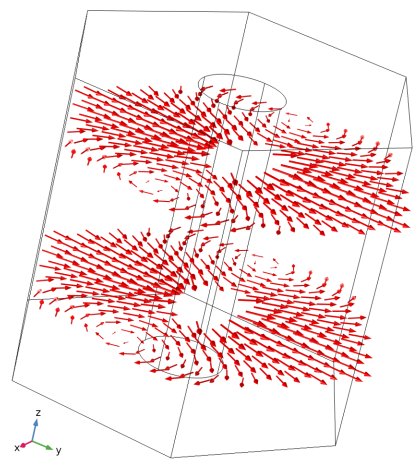}\label{fig:FieldLinesMode1}}
		\sidesubfloat[]{\includegraphics[width=0.2\columnwidth]{CylinderAntiSymmetricMode.png}\label{fig:FieldLinesMode2}}
		
		\smallskip
		\sidesubfloat[]{\includegraphics[width=0.2\columnwidth]{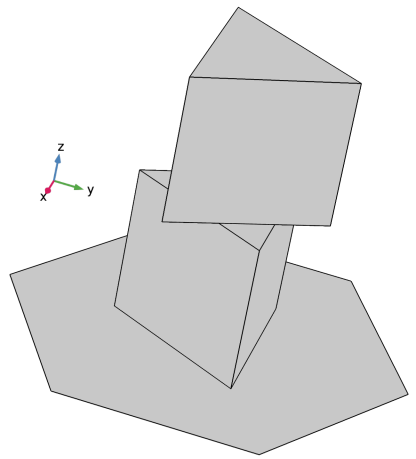}\label{fig:doubleTripodsPhC}}
		\sidesubfloat[]{\includegraphics[width=0.2\columnwidth]{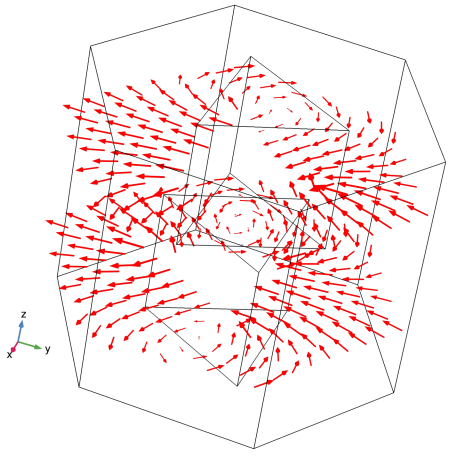}\label{fig:D_FieldLinesMode1}}
		\sidesubfloat[]{\includegraphics[width=0.2\columnwidth]{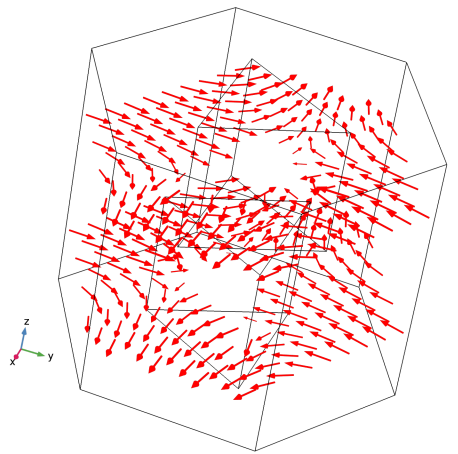}\label{fig:D_FieldLinesMode2}}
		
		\label{fig:unperturbedFields}
		\caption{\textbf{(a)} The unperturbed PhC unit cell. \textbf{(b)} The field lines of the in-plane electric field at $z=0.25a_0$ and $z=0.75a_0$ for one of the antisymmetric modes of the unperturbed cylinder unit cell. \textbf{(c)} Same as (b) but for a symmetric mode. \textbf{(d)} The perturbed non-rotated double triangle unit cell. \textbf{(e)} The field lines of the in-plane electric field for one of the antisymmetric modes of the non-rotated double triangle unit cell. Same as (e) but for a symmetric mode.}
	\end{figure}

	We first examine the properties of a system consisting of two perfect electric conductor (PEC) plates with periodic PEC cylinders placed in a triangular lattice between them, depicted in Fig. \ref{fig:unperturbedPhC}. Due to the triangular lattice the system exhibits two Dirac cones at the $K$ and $K'$ points of the reciprocal lattice. By judiciously choosing the distance between the plates, the radius of the cylinders and the size of the gaps separating them from the plates it was shown that the two Dirac cones can become degenerate. Due to the mirror symmetry of our system regarding the mid-plane, we can define TE-like and TM-like modes according to their symmetry properties and label each mode accordingly \cite{joannopoulos2011photonic}. 
	
	Since we intend to break the mid-plane mirror symmetry, we choose to work with a more convenient degree of freedom for the structure that we discuss. We define the a symmetric and antisymmetric mode as even and odd in the in-plane electric field regarding an out-of plane flip respectively. This symmetry is depicted in Fig. \ref{fig:FieldLinesMode1} and \ref{fig:FieldLinesMode2}. One can see that the unit-cell is invariant to $180^\circ$ rotation around an axis at $z=h/2$ normal to the $xz$ plane, resulting in the in-plane electric field being either symmetric or anti-symmetric with regards to that operation. By virtue of the triangular lattice, our system possesses an additional degree of freedom which is the right-hand circular polarization (RCP) and left-hand circular polarization (LCP), which can always be constructed as a linear combination of the two degenerate symmetric and antisymmetric modes at the $K$ or $K'$ points separately. The RCP and LCP modes are presented in Fig. \ref{fig:unperturbedModes}. In Fig. \ref{fig:doubleTripodsPhC} we present the double triangle unit cell with $\alpha = 0$ and show in Fig. \ref{fig:D_FieldLinesMode1} and \ref{fig:D_FieldLinesMode2} that this structure maintains the symmetry to $180^\circ$ rotation around an axis at $z = h/2$ normal to the $xz$ plane, resulting in perturbed modes that are still symmetric and antisymmetric in the in-plane $E$ field.
	
	\section{Derivation of the photonic graphene Hamiltonian}
	Because only two symmetric and antisymmetric modes exist in the $K$ and $K'$ points, any electromagnetic mode with Bloch wavenumber $\mathbf{k}$ can be expanded as 
	\begin{equation}
	\begin{split}
	\mathbf{E}\left(\mathbf{r},t \right) = \sum_{n,\mathbf{k}}\left[ a_{s}^{n}(\mathbf{k})\mathbf{e}_{s}^{n,\mathbf{k}}\left( \mathbf{r}_{\perp},z\right) +  a_{a}^{n}(\mathbf{k})\mathbf{e}_{a}^{n,\mathbf{k}}\left( \mathbf{r}_{\perp},z\right) \right] \\  \times e^{i\mathbf{k}\cdot{}\mathbf{r_\perp}-i\omega_n(\mathbf{k})t} + \textup{c.c.}
	\end{split}
	\end{equation}
	
	\begin{equation}
	\begin{split}
	\mathbf{H}\left(\mathbf{r},t \right) = \sum_{n,\mathbf{k}}\left[ a_{s}^{n}(\mathbf{k})\mathbf{h}_{s}^{n,\mathbf{k}}\left( \mathbf{r}_{\perp},z\right) +  a_{a}^{n}(\mathbf{k})\mathbf{h}_{a}^{n,\mathbf{k}}\left( \mathbf{r}_{\perp},z\right) \right] \\  \times e^{i\mathbf{k}\cdot{}\mathbf{r_\perp}-i\omega_n(\mathbf{k})t} + \textup{c.c.,} 
	\end{split}
	\end{equation}
	where the $n=1,2$ refers to the lower and upper propagation bands and $\mathbf{e}_{s}^{n,\mathbf{k}},\mathbf{h}_{s}^{n,\mathbf{k}},\mathbf{e}_{a}^{n,\mathbf{k}},\mathbf{h}_{a}^{n,\mathbf{k}}$ are the normalized periodic field profiles in the $\mathbf{r}_\perp = (x,y)$ plane, and $s(a)$ labels the symmetric (antisymmetric) modes. We obtain the dispersion relation and field profiles by using COMSOL Multiphysics to solve for the eigenmodes of the unperturbed PhC with Bloch-Floquet periodic boundary conditions and by carefully tuning the geometric parameters of the PhC we achieve degeneracy for the two Dirac cones, giving four degenerate modes that are spanned by the symmetric/antisymmetric and RCP/LCP polarizations at the $K/K'$ points. The effective Hamiltonian around the $K$ valley for the symmetric(antisymmetric) polarization is then $H_{0,K}^{s(a)} = \nu_{D}\left( \delta{}k_x\hat{\sigma}_x+\delta{}k_y\hat{\sigma}_y\right)$, where $\hat{\sigma}_{x,y,z}$ are the Pauli matrices acting on the RCP/LCP state vector $\mathbf{U}_{K}^{s(a)} = \left[ a_{s(a)}^R ; a_{s(a)}^L\right].$ The basis $\mathbf{U}_R = [1;0]$ and $\mathbf{U}_L = [0;1]$ is defined according to its transformation with respect to $120^\circ$ rotation $\mathcal{R}_3$ according to $\mathcal{R}_3{}\mathbf{U}_{R,L} = \exp{\left( \mp2\pi i /3\right) }\mathbf{U}_{R,L}$. 
	
	Using the degeneracy of the symmetric and antisymmetric modes we can expand the basis states separate two component vectors $\mathbf{U}_{K}^{s},\mathbf{U}_{K}^{a}$ to a single four component vector $\mathbf{V}_{K} = \mathbf{M}\left[ \mathbf{U}_{K}^{s} ; \mathbf{U}_{K}^{a} \right] $, where $\mathbf{M}$ is an arbitrary unitary $4\times 4$ matrix that does not mix the RCP and LCP polarizations but couples the symmetric/antisymmetric states. We choose to work with $\mathbf{M} = \mathbf{I} $, the identity matrix, for reasons that will becomes clear once we introduce perturbations to our unit cell structure. Combining the symmetric and antisymmetric modes to a four component vector allows us to write the combined symmetric and antisymmetric effective Hamiltonian as $H_{0,K} = \nu_{D}\hat{s}_{0}\left( \delta{}k_x\hat{\sigma}_x+\delta{}k_y\hat{\sigma}_y\right)$, where $\hat{s}_{x,y,z}$ are the symmetric and antisymmetric modes' basis Pauli matrices with $\hat{s}_0$ the $2\times2$ identity matrix in this basis. A Kronecker product is implied between the Pauli matrices of the RCP/LCP and the symmetric/antisymmetric bases. We note that since we have a symmetric mode and an anti-symmetric mode with this choice of basis (see preceding section), performing a $C'_2$ out-of-plane flip operation in the $\hat{y}$ direction is identical to operating with $\hat{s}_z$ along with $k_y\rightarrow -k_y$ on the Hamiltonian; we will return to this later when we discuss the symmetries of the unit cell. The final basis expansion is obtained by including both the $K$ and $K'$ valleys of the Brillouin Zone (BZ) by introducing an eight component spinor, $\mathbf{\Psi} = \left[ \mathbf{V}_K ; \mathbf{TV}_{K'} \right] $, where the transformation matrix $\mathbf{T}$ swaps the RCP and LCP polarization states. By introducing the Pauli matrices $\hat{\tau}_{x,y,z}$ and $\hat{\tau}_0$ operating on the valley subspace and using symmetry considerations, the effective unperturbed $8 \times 8$ Hamiltonian describing the PhC can be written as $H_{0} = \nu_{D}\left( \delta{}k_x\hat{\tau}_z\hat{s}_{0}\hat{\sigma}_x+\delta{}k_y\hat{\tau}_0\hat{s}_{0}\hat{\sigma}_y\right)$, which is exactly the low-energy Hamiltonian of graphene. 
	
	\section{Slater cavity perturbation theory}
	We now introduce Slater cavity perturbation theory \cite{S_slater1946microwave} and its generalization for degenerate unperturbed modes \cite{S_dombrowski1984matrix} to derive matrix expressions in the valley, symmetric/antisymmetric and RCP/LCP basis for gap opening terms in the effective Hamiltonian. We only consider perturbations that alter the shape of the PEC boundaries of the unit cell of the PhC.
	
	The perturbed electric and magnetic fields at $\delta\mathbf{k} = 0$ can be written as a linear combination of the unperturbed electric and magnetic fields respectively: 
	\begin{equation}
	\mathbf{E}'=\sum_n a_n \mathbf{E}_n \textup{ and } \mathbf{H}' = \sum_n a_n \mathbf{H}_n,
	\end{equation}
	where $\left( \mathbf{E}_n, \mathbf{H}_n \right) $ are the unperturbed fields at $\delta \mathbf{k} = 0$ and the coefficients $a_n$ describe the hybridization between them. The generalized Slater matrix system is then
	\begin{equation}
	\begin{pmatrix}
	\omega_{11} + \kappa_{11} & \kappa_{12} & \dots \\
	\kappa_{21} &    \omega_{22} + \kappa_{22}  & \dots \\
	\vdots & \vdots  & \ddots
	\end{pmatrix}
	\begin{pmatrix}
	a_1 \\
	a_2 \\
	\vdots
	\end{pmatrix}
	=
	\omega'
	\begin{pmatrix}
	a_1\\
	a_2\\
	\vdots
	\end{pmatrix},
	\label{eq:CMT}
	\end{equation}
	where $\omega_{mn}$ with $m,n = 1,2,3\dots,$ is the eigenfrequency of the unperturbed modes, and $\omega'$ is the frequency of the perturbed modes which is obtained by solving the eigenvalues problem of Eq. (\ref{eq:CMT}), which defines the effective Hamiltonian for the perturbed PhC as it is analogous to the time independent Schr\"{o}dinger equation. The coupling coefficients $\kappa_{mn}$ are given by the overlap integral: 
	\begin{equation}
	\kappa_{mn} = -\int_{\Delta V}\left( \omega_m \mathbf{E}^{*}_m\cdot\mathbf{E}_n - \omega_n\mathbf{H}^{*}_m\cdot\mathbf{H}_n\right)dV, 
	\end{equation}
	where $\Delta V$ is the perturbed volume where an extra piece of metal is inserted and the eigenfields are normalized as $\int_V \left( \left| \mathbf{E}_n \right|^2 + \left|\mathbf{H}_n \right|^2  \right)dV  = 1$, with $V$ being the volume of the unperturbed unit cell. Due to the degeneracy of our modes we can define a dimensionless coupling strength: 
	\begin{equation}
	\Delta_{mn} \equiv \kappa_{mn}/\omega_D = -\int_{\Delta V}\left( \mathbf{E}^{*}_m\cdot\mathbf{E}_n - \mathbf{H}^{*}_m\cdot\mathbf{H}_n\right)dV. 
	\label{eq: matEl}
	\end{equation}
	For the case of $m=n$ the integrand is the Lagrangian density of the electromagnetic fields \cite{S_jackson2012classical}. As in the main text, we choose the following notation for the symmetric/antisymmetric and RCP/LCP basis 
	\begin{equation}
	\begin{pmatrix}
	a_1\\
	a_2\\
	a_3\\
	a_4
	\end{pmatrix} = \begin{pmatrix}
	S,RCP \\
	S,LCP \\
	A, RCP\\
	A,LCP
	\end{pmatrix}, \label{eq:basis}
	\end{equation}
	i.e. $\Delta_{13}$ couples the symmetric RCP mode to the antisymmetric RCP mode, $\Delta_{22}$ couples the symmetric LCP mode to itself, etc. 
	
	\section{Non-rotated double triangle perturbation}
	The first perturbation we apply this formalism to is the double triangle perturbation that is not rotated w.r.t. the hexagonal unit cell, meaning $\alpha = 0^\circ$. This perturbation conserves the out-of-plane flip symmetry which defined the symmetric and antisymmetric basis. Therefore, it will affect the symmetric and antisymmetric subspaces separately and will not couple them, meaning $\Delta_{mn}=0$ for the off-diagonal blocks. Furthermore, due to the $C_3$ symmetry of the perturbation, the perturbation is diagonal in the symmetric/antisymmetric and RCP/LCP basis, meaning $\Delta_{mn} = 0$ for $m\neq n$ in the diagonal blocks. To prove that we inspect Eq. (\ref{eq: matEl}) and note that due to $C_3$ symmetry \begin{equation}
	\Delta_{RL}^{s(a)} = \mathcal{R}_3\Delta_{RL}^{s(a)}.
	\end{equation}
	From the definition of the RCP and LCP polarizations we know that $\mathcal{R}_3\mathbf{e}_{s(a)}^R = \eta^{*}\mathbf{e}_{s(a)}^R$ and $\mathcal{R}_3\mathbf{e}_{s(a)}^L = \eta\mathbf{e}_{s(a)}^L$, where $\eta = \exp(i2\pi/3)$. Plugging those relations in Eq. (\ref{eq: matEl}) we find 
	\begin{equation}
	\begin{split}
	\Delta_{RL}^{s(a)} = & \int_{\Delta V} \left( \mathcal{R}_3\mathbf{e}_{s(a)}^R\right)^*\cdot\left( \mathcal{R}_3\mathbf{e}_{s(a)}^L\right) - \\  & \left(\mathcal{R}_3\mathbf{h}_{s(a)}^R\right)^*\cdot\left(\mathcal{R}_3\mathbf{h}_{s(a)}^L \right) dV \\
	= &  \eta^2 \int_{\Delta V}\left( \mathbf{e}_{s(a)}^R\right)^*\cdot \mathbf{e}_{s(a)}^L - \left( \mathbf{h}_{s(a)}^R\right)^*\cdot \mathbf{h}_{s(a)}^L dV\\
	=  &  \eta^2\Delta_{RL}^{s(a)}, 
	\label{eq:RCPLCP}
	\end{split}
	\end{equation}
	which means $\Delta_{RL}^{s(a)} = 0$.
	
\begin{figure*}
		\centering
		\sidesubfloat[]{\includegraphics[width=0.2\paperwidth]{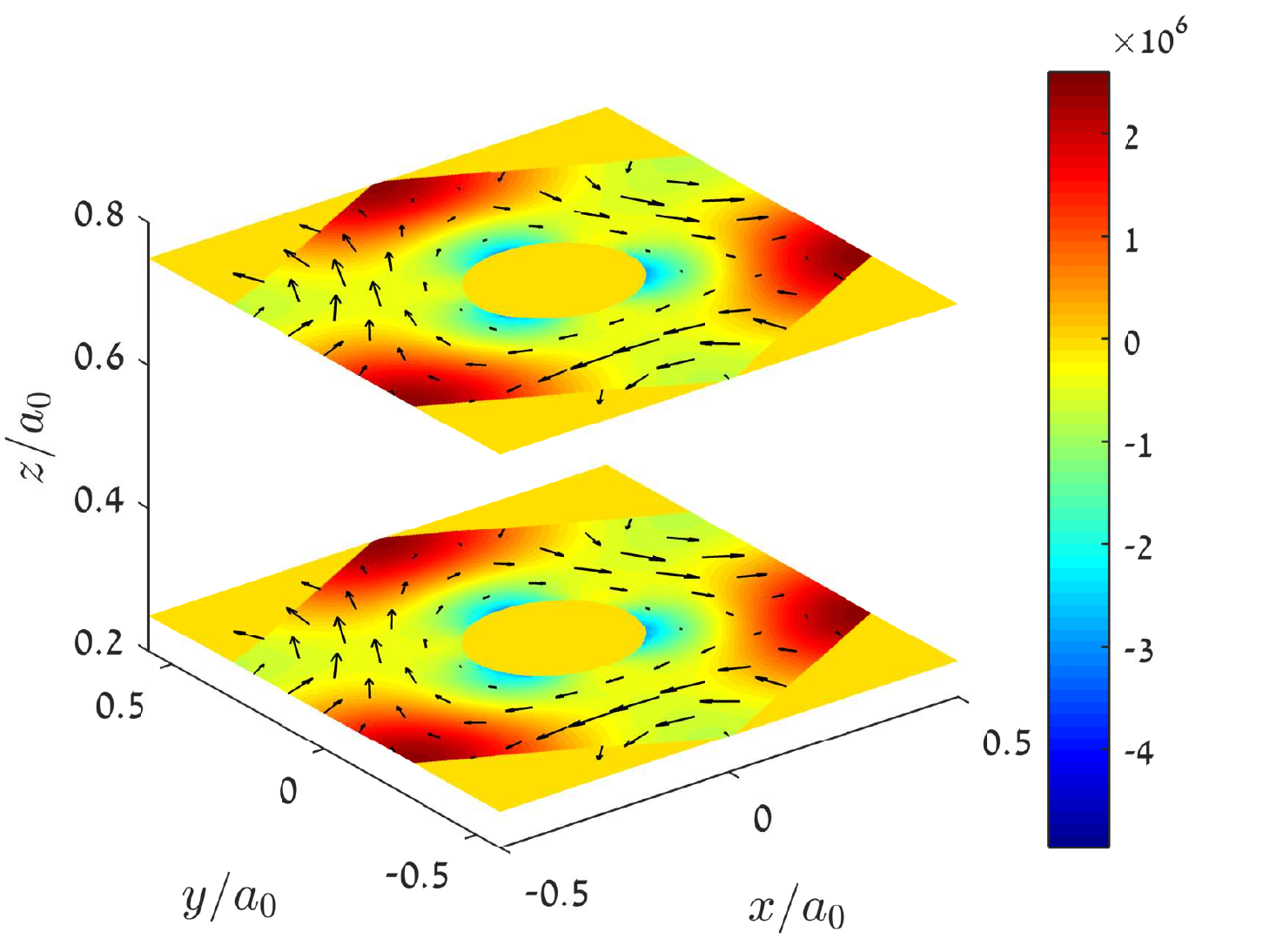}\label{fig:TM_LCP}}
		\sidesubfloat[]{\includegraphics[width=0.2\paperwidth]{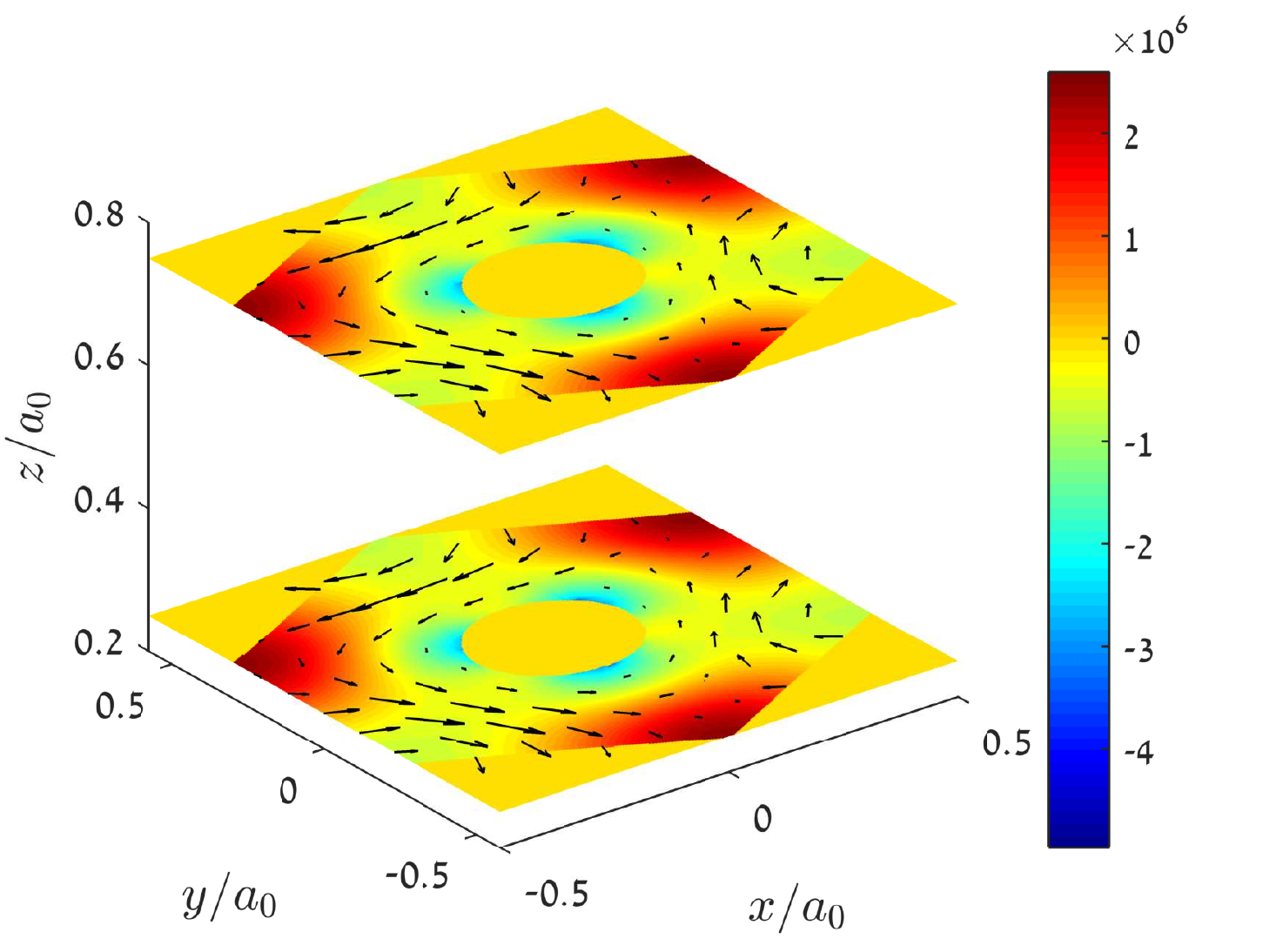}\label{fig:TM_RCP}}
		\sidesubfloat[]{\includegraphics[width=0.2\paperwidth]{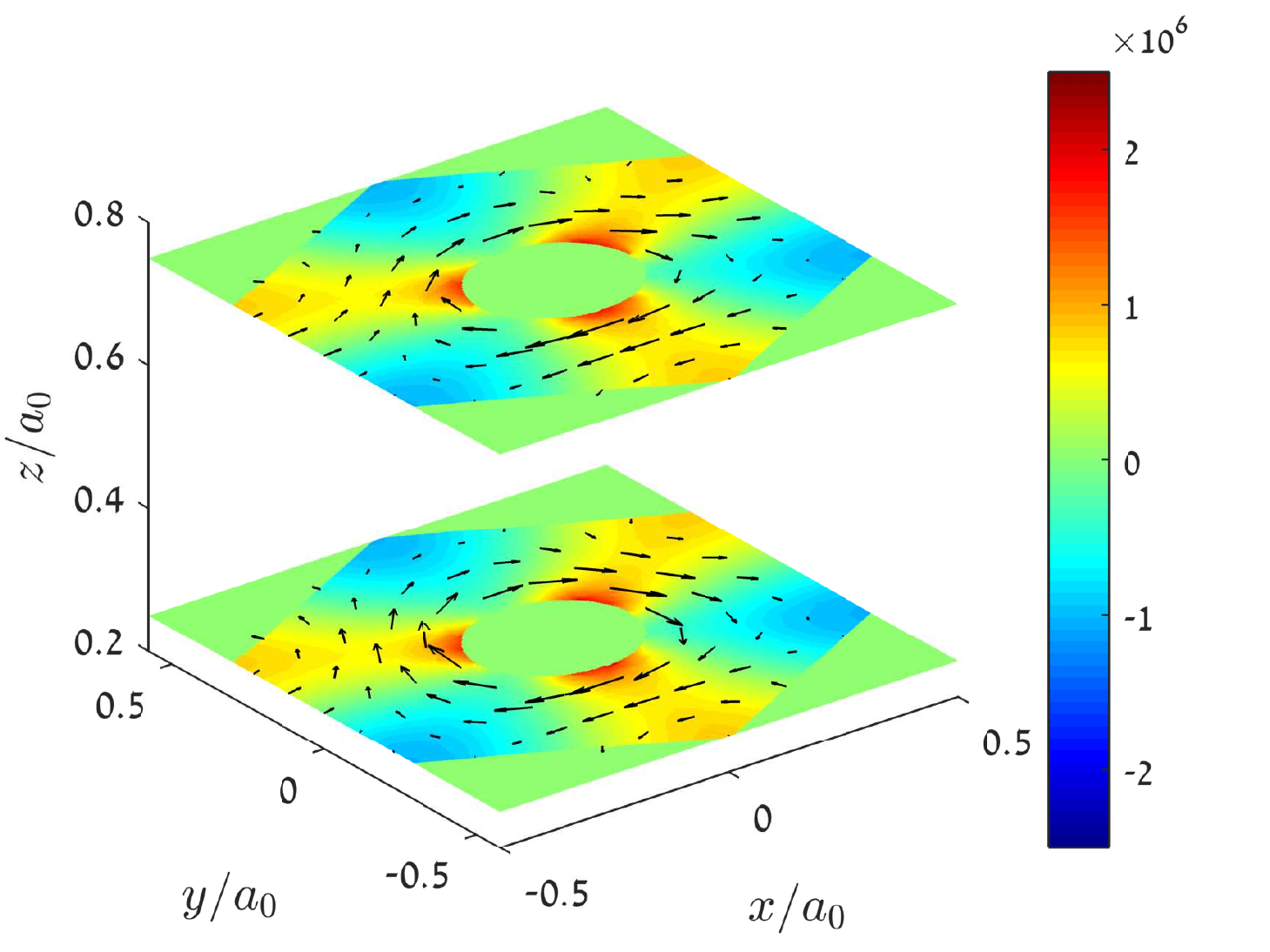}\label{fig:TE_LCP}}
		\sidesubfloat[]{\includegraphics[width=0.2\paperwidth]{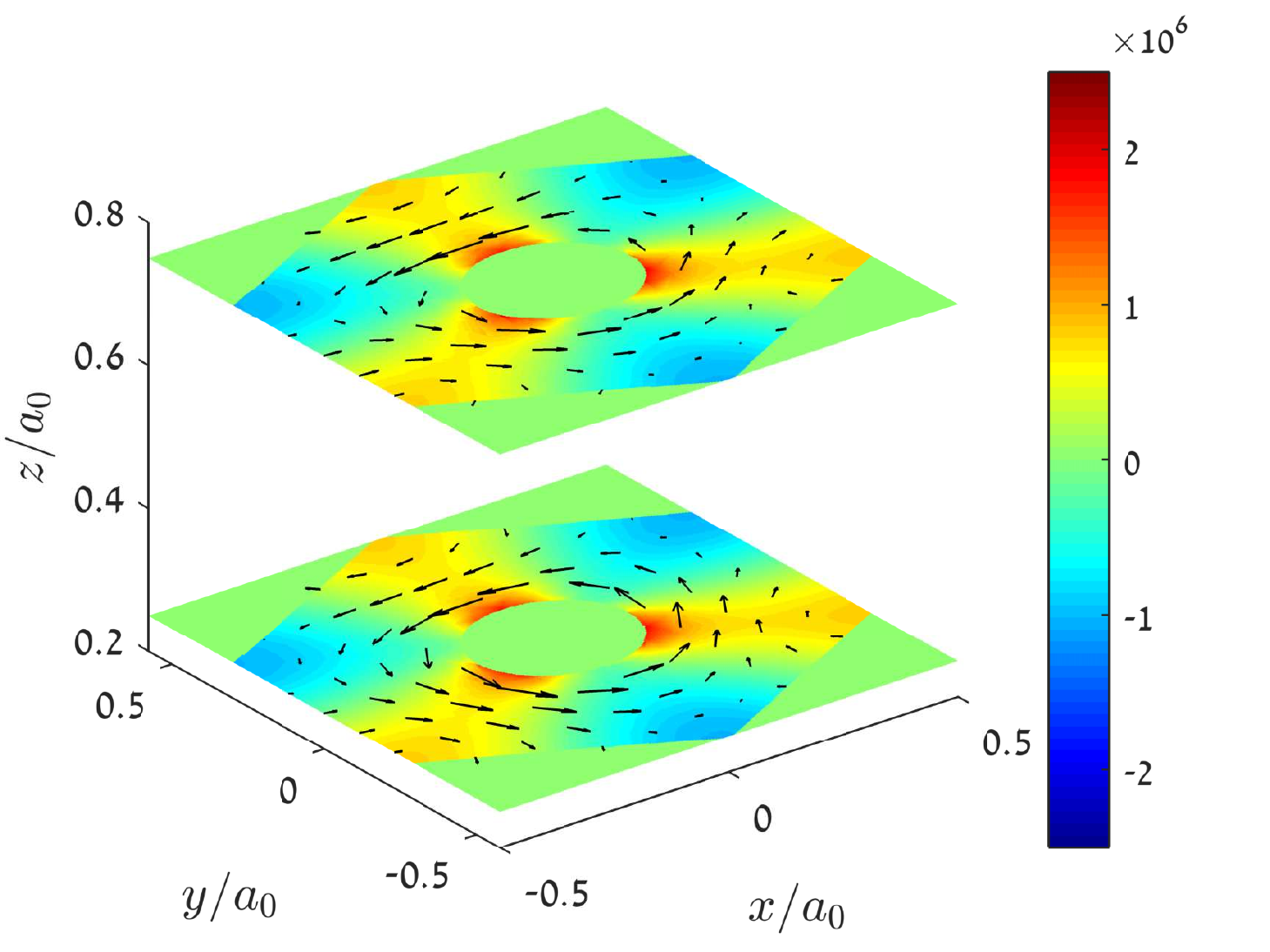}\label{fig:TE_RCP}}
		\caption{Lagrangian densities in color and time averaged power flux in black arrows at $z=h/4$ and $z=3h/4$ for the unperturbed PhC for \textbf{(a)} antisymmetric LCP mode \textbf{(b)} antisymmetric RCP mode \textbf{(d)} symmetric LCP mode and \textbf{(d)} symmetric RCP mode.}
		\label{fig:unperturbedModes}
	\end{figure*}

	Combining the last two results we reach the conclusion that $\Delta_{mn} = 0$ for $m\neq n$, meaning the non-rotated double triangle perturbation is diagonal in the RCP/LCP symmetric/antisymmetric basis. To see that the perturbation takes the form $H_D = \Delta_D\hat{s}_z\hat{\sigma}_0$ we inspect Eq. (\ref{eq: matEl}) for $m = n$ plotted in Fig. \ref{fig:unperturbedModes} and note that $\Delta V$, the volume displaced by the non-rotated double triangle metallic perturbation, displaces the same total Lagrangian density for the RCP and LCP states of the symmetric and the antisymmetric modes respectively. This is the reason why the non-rotated double triangle perturbation does not gap the respective symmetric and antisymmetric Dirac cones, but rather shifts them in frequency, thereby removing their degeneracy. The non-rotated double triangle perturbation $\Delta_D$ at this point is simply the frequency difference between the respective Dirac cones, where a shared shift in frequency is possible too. Therefore $\Delta_D = (\Delta_{RL}^{s}-\Delta_{RL}^{a})/2$ and there is a shared shift in frequency of both Dirac cones by a frequency of $(\Delta_{RL}^{s}+\Delta_{RL}^{a})/2$.
	
	To add the valley degree of freedom we note that switching valleys is equivalent to switching the RCP and LCP modes forming the Dirac cones, but since the double non-rotated double triangle perturbation does not distinguish between the RCP and LCP modes, there is no effect due to switching from the $K$ point to $K'$. Hence the full perturbation term for the double non-rotated double triangles is $H_D = \Delta_D\hat{\tau}_0\hat{s}_z\hat{\sigma}_0$. We will show a more rigorous proof for the form of the valley degree of freedom in a proceeding section of the supplementary.
	
	\section{Rotated double triangles perturbation}
	Next we derive the form of the rotated double triangles perturbation term of the effective Hamiltonian. The rotated perturbation breaks the out-of-plane flip $C'_2$ symmetry and therefore couples the symmetric and antisymmetric subspaces, which in turn means it is not a diagonal perturbation. Since this perturbation maintains $C_3$ symmetry which defines the RCP and LCP basis, we do not get any coupling between LCP and RCP modes at the valleys, which is easily proven using Eq. (\ref{eq:RCPLCP}) if we allow for one mode to be symmetric and the other antisymmetric. This means that the coupling matrix has non-zero elements only for terms that couple RCP to RCP and LCP to LCP, i.e. $\Delta_{13},\Delta_{24},\Delta_{31},\Delta_{42}$ and the diagonal elements. Because of the RCP and LCP degeneracy of the Lagrangian density with regards to the rotated double triangle perturbation, the diagonal elemenets take the same form as they did for the non-rotated double triangles, which is the reason that the Dirac cones do not gap for the rotated perturbation either. We can therefore effectively treat the diagonal terms of the rotated perturbation as part of the non-rotated double triangles perturbation and from here on only treat the off-diagonal blocks.
	
	We already concluded that the coupling matrix is composed of only off-diagonal blocks in which only the diagonal elements are non-vanishing. This forces the matrix describing the symmetric and antisymmetric coupling to be proportional to a linear combination of $\hat{s}_x$ and $\hat{s}_y$. Without loss of generality we will assume our effective Hamiltonian only has the component proportional to $\hat{s}_x$, otherwise we can always rotate around the $\hat{s}_z$ axis such that the rotated double triangle term in the effective Hamiltonian is proportional only to $\hat{s}_x$. To determine the RCP/LCP Pauli matrix we note that it can only be $\hat{\sigma}_z$ or $
	\hat{\sigma}_0$ because otherwise the off-diagonal blocks are not diagonal. However, if we choose $\hat{\sigma}_0$ we can just rotate $\hat{s}$ to the direction given by $\Delta_D\hat{s}_z + \Delta_\alpha \hat{s}_x$ to get a combined term proportional to $\hat{s}_z\hat{\sigma}_0$ which is equivalent to the non-rotated double triangle term alone.  This means that the rotated double triangles perturbation effective Hamiltonian term is $H_\alpha = \Delta_\alpha\hat{s}_x\hat{\sigma}_z$, where $\Delta_{\alpha} = \Delta_{13}=-\Delta_{24}$ after the proper unitary transformation on the $\hat{s}$ DOF described above has been performed. 
	
	\section{Inclusion of the valley degree of freedom}
\begin{figure}[b]
		\centering
		\sidesubfloat[]{\includegraphics[width=0.3\columnwidth]{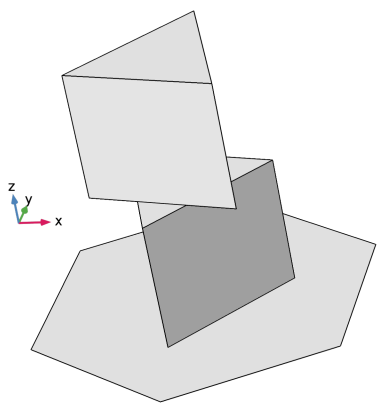}\label{fig:UnitCellPers}}
		\sidesubfloat[]{\includegraphics[width=0.6\columnwidth]{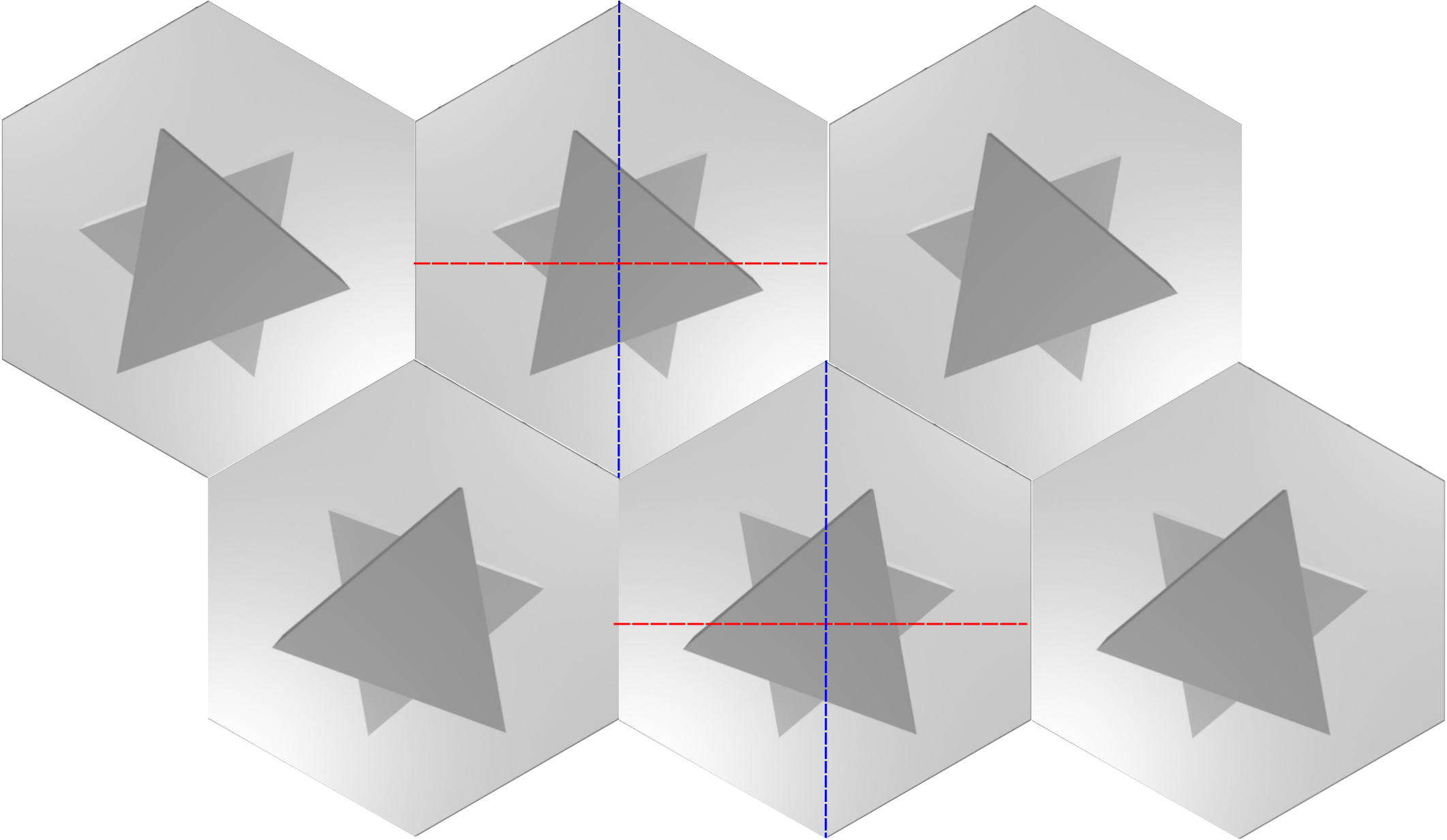}\label{fig:unitCell1}}
		\caption{\textbf{(a)} Perspective view of the double triangles unit cell with $\alpha \neq 0^\circ$. \textbf{(b)} Top view of the rotated double triangles interface region between two different domains. The blue dashed lines marks the $\hat{y}$ direction of the different unit cells and the reflection plane used for the reflection operator $\mathcal{R}$. The red dashed lines marks the axis around which we perform the $C'_2$ out-of-plane flip.}
		\label{fig:RefInv}
	\end{figure}

	We add the valley Pauli matrix by demanding that our Hamiltonian has an inversion operator $\mathcal{P}$ that obeys the inversion relation \cite{asboth2016short}
	\begin{equation}
	\mathcal{P}H_K(q_x,q_y)\mathcal{P}^{-1} = H_{K'}(-q_x,-q_y), 
	\label{eq:Inv}
	\end{equation}
	where
	\begin{equation}
	H_K(q_x,q_y) = \nu_{D}\left( q_{x}\hat{s}_{0}\hat{\sigma}_{x} + q_{y}\hat{s}_{0}\hat{\sigma}_{y} \right) + \Delta_D \hat{s}_{z}\hat{\sigma}_{0} +\Delta_{\alpha}\hat{s}_{x}\hat{\sigma}_{z}
	\label{eq:ValleyHamiltonian}
	\end{equation}
	and $H_{K'}(q_x,q_y)$ is identical to $H_K(q_x,q_y)$ up to possible sign changes of the different terms, implying an assumption of no valley mixing. We set to find $\mathcal{P}$ and in order to do that we notice that an inversion operator can be composed of a reflection operator and an out of plane flip which together leave the unit cell invariant, as can be seen by inspecting Fig. (\ref{fig:RefInv}).
	
	Using the definition of the RCP and LCP states we define the reflection operator as  $\mathcal{R} = \hat{s}_0\hat{\sigma}_x$ along with $k_x\rightarrow-k_x$ and using the definition of the symmetric and anti-symmetric modes we define the out-of-plane flip operator as  $\mathcal{F} = \hat{s}_z\hat{\sigma}_0$ along with $k_y\rightarrow-k_y$ (see Fig. \ref{fig:unitCell1}). Composing the two operators results in an inversion operator
	\begin{equation}
	\mathcal{P} = \hat{s}_z \hat{\sigma}_x,
	\end{equation}
	that must obey the inversion relation of Eq. (\ref{eq:Inv}). Applying Eq. (\ref{eq:Inv}) we find the Hamiltonian at $K'$ which is identical to the Hamiltonian of Eq. (\ref{eq:ValleyHamiltonian}) up to a change of sign for the term $ q_{x}\hat{s}_{0}\hat{\sigma}_{x}$. We therefore find the full Hamiltonian combining both valleys:
	\begin{equation}
	H = \nu_{D}\left( q_{x}\hat{\tau}_{z}\hat{s}_{0}\hat{\sigma}_{x} + q_{y}\hat{\tau}_{0}\hat{s}_{0}\hat{\sigma}_{y} \right) + \Delta_D\hat{\tau}_{0} \hat{s}_{z}\hat{\sigma}_{0} +\Delta_{\alpha}\hat{\tau}_{0}\hat{s}_{x}\hat{\sigma}_{z}.
	\end{equation}
	Of course we can easily verify that for the tight binding model the relation 
	\begin{equation}
	\mathcal{P}H(k_x,k_y)\mathcal{P}^{-1} = H(-k_x,-k_y)
	\end{equation}
	holds.
	
	\section{Proof that the sign of $\Delta_\alpha$ flips between domains}
	For the supercell simulations in the main text we use two interfacing PhCs that are distinguished from each other by either a reflection in the $\hat{x}$ direction or a $C'_2$ out-of-plane flip operation. Here we prove that the effect of either operations on the unit cell of a domain changes the sign of $\Delta_\alpha$. Applying the reflection operator $\mathcal{R}$ defined in the preceding section we find the relation 
	\begin{equation}
	\mathcal{R}H^{(1)}(-k_x,k_y)\mathcal{R}^{-1} = H^{(2)}(k_x,k_y), 
	\end{equation}
	where, $H^{(1)}$ is the Hamiltonian of Eq. (2) of the main text and $H^{(2)}$ is the same Hamiltonian but with $\Delta_\alpha\rightarrow-\Delta_\alpha$ and $f_{\mathbf{k}}\rightarrow f^{*}_\mathbf{k}$. 
	Since $f_\mathbf{k} = 1+2\cos(k_x a_0)\exp{(i\sqrt{3}k_y a_0)}$ a reflection in $x$ transforms $k_x$ to $-k_x$ has no effect due to the cosine function being even, but the valley expansion position changes valleys from  $\mathbf{K} = (4\pi/3a_0,0)$ to $\mathbf{K}' = (-4\pi/3a_0,0)$ and vice versa. We therefore find an overall minus sign to the expansion of $f_\mathbf{k}$ which is inconsequential to the existence of edge states, confirming that this indeed does not flip the valley degree of freedom.
	
	Alternatively, one can choose to operate with the $C'_2$ operator $\mathcal{F} = \hat{s}_z\hat{\sigma}_0$ along with $k_y\rightarrow-k_y$ instead of reflection. The obtained relation is then 
	\begin{equation}
	\mathcal{F}H^{(1)}(k_x,-k_y)\mathcal{F}^{-1} = H^{(2)}(k_x,k_y). 
	\end{equation}
	Here, the change in sign of $k_y$ conjugates $f_\mathbf{k}$. We show a rigorous proof that the phase of $f_\mathbf{k}$ plays no role in determining the topology of the Hamiltonian in the mapping the tight binding model to AFM bilayer graphene section below.
	
	\section{Calculation of kink states spectra using the tight binding model}
	In this section we describe the structure of the tight binding supercell simulations in more detail. We model two domains using the tight binding Hamiltonian of Eq. (2) of the main text with alternating sign of $\Delta_{\alpha}$ between domains. To do that, we assume translation symmetry in the $\hat{x}$ direction and a  $N=50$ unit cells in the $\hat{y}$ and $-\hat{y}$ direction with closed boundary conditions, meaning we assume unit cell $1$ and $2N$ are neighbors. The basis vector is therefore now 
	\begin{equation}
	\psi_{k_x} =
	\begin{pmatrix}
	\begin{pmatrix}
	S,RCP \\
	S,LCP \\
	A,RCP \\
	A,LCP
	\end{pmatrix}_1
	\\
	\begin{pmatrix}
	S,RCP \\
	S,LCP \\
	A,RCP \\
	A,LCP
	\end{pmatrix}_2
	\\
	\vdots
	\\
	\begin{pmatrix}
	S,RCP \\
	S,LCP \\
	A,RCP \\
	A,LCP
	\end{pmatrix}_{2N}
	\end{pmatrix},
	\end{equation}
	where the sign of $\Delta_{\alpha}$ switches sign between unit cells $1\dots
	N$ and $N+1\dots2N$ and $\Delta_D$ remains unchanged. The transition amplitudes between neighboring unit cells in the supercell Hamiltonian are $-t(1+\exp(2\pi i k_x a_0))$ for unit cells $1\dots N$ and $-t(1+\exp(-2\pi i k_x a_0))$ for unit cells $N+1\dots 2N$, where we choose $t=1$ and the transition amplitudes connecting the different domains was chosen without conjugation. The supercell Hamiltonian is a $2N\times2N$ matrix made of $4\times4$ blocks with transition amplitudes connecting the different $4\times4$ unit cell blocks. After constructing the supercell Hamiltonian matrix, we numerically diagonalize it for every $k_x$ value to obtain the edge state spectrum. This results in two edge states on the interface between domains as well as identical edge states at the termination in the $\pm\hat{y}$ direction. Note that we have verified that conjugation of the transition amplitudes connecting the different domains does not change the edge state spectrum.
	
	\section{Calculation of edge mode spectra according to the Dirac theory}
\begin{figure}
		\centering
		\includegraphics[width=0.3\columnwidth]{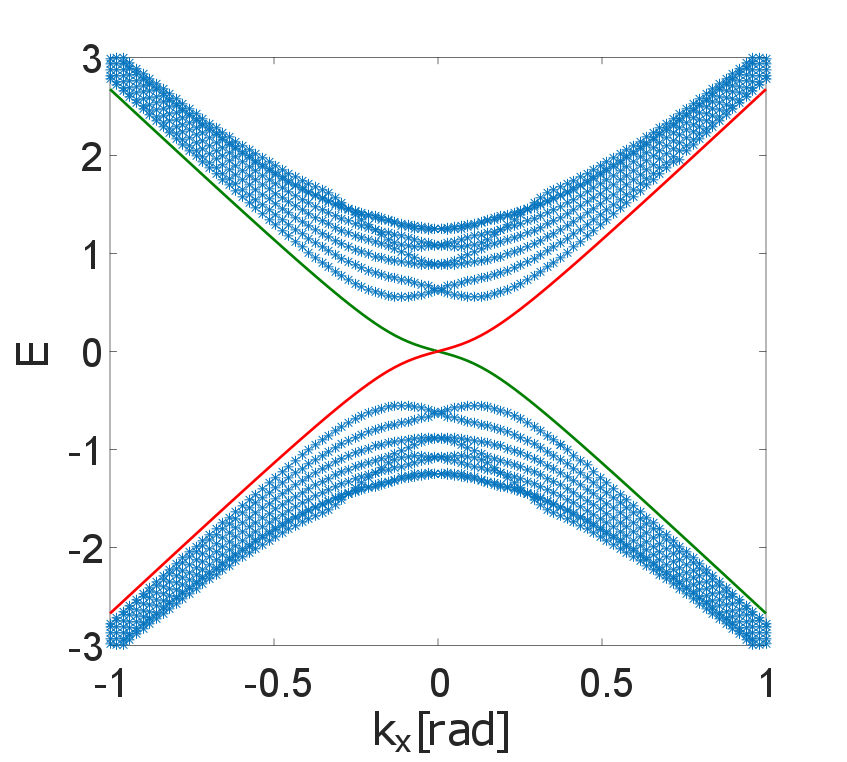}
		\caption{Edge spectrum obtained from solving the eigenvalues of Eq. (\ref{eq:ValleyHamiltonian}). Bulk modes are marked with blue asterisks, edge modes are marked in red and green solid lines. The values of the parameters used are $\nu_{D} = 1, \Delta_D = 0.5,  \Delta_\alpha = 4, a = 100. $ }
		\label{fig:DiracTheorySpectrum}
	\end{figure}

	In this section we show that the Hamiltonian found from first principles at the $K$ point Eq. (\ref{eq:ValleyHamiltonian}) agrees with the tight binding model results for small $\Delta_D$ values. We use COMSOL Multiphysics math module for 1D geometry to solve the equation $\hat{H}\mathbf{u} = E\mathbf{u}$, where $\mathbf{u}$ is an eigenvector of $\hat{H}$ and $E$ is the eigenvalue, on a line of length $a$. We assume periodicity in the $x$ direction, hence $k_x$ is a conserved quantity, but we use open boundary conditions at the terminations of the domain in the $\pm\hat{y}$ directions. We therefore use the substitution $k_y\rightarrow -i\partial_y$ and write the eigenvalue equation in coefficient form. To model the change in sign of $\Delta_\alpha$ between two domains at the domain wall, we use a position dependent value for $\Delta_\alpha$ such that $\Delta_\alpha\rightarrow\Delta_\alpha\tanh{\left( y/b \right) }$ where $a$ is a scaling constant controlling the rate of the transition from $\Delta_\alpha$ to $-\Delta_\alpha$. The resulting spectrum of bulk and edge states is shown in Fig. \ref{fig:DiracTheorySpectrum} . One can observe the effect of the gapping of the nodal line on the bulk states by noting split maximas and minimas of the bulk states. We can also identify the change in group velocity of the edge modes due to the presence of non-zero $\Delta_D$. 
	
	\section{Mapping the low energy model to graphene with AFM order}
\begin{figure}
		\centering
		\includegraphics[width=1\columnwidth]{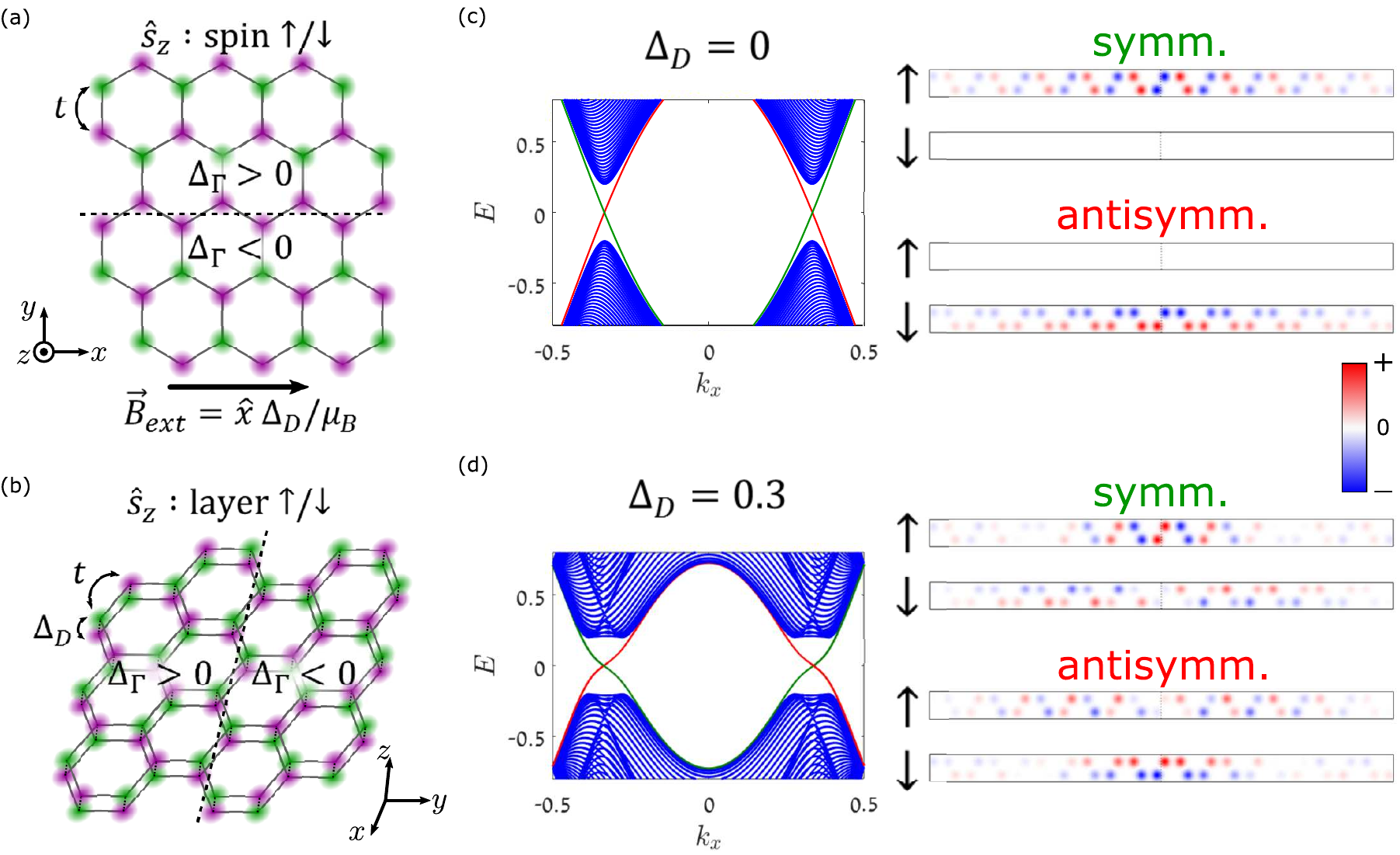}
		\caption{(a) Visualization of mapping between our photonic crystal and graphene with AFM order under an external magnetic field. The sign of $\Delta_\alpha$ determines the ordering of the different spins, in green and purple, in each domain. (b) Visualization of mapping between our photonic crystal and AA-stacked bilayer graphene. The sign of $\Delta_\alpha$ determines the ordering of different spins in each domain as in the single layer case. Note that the intrinsic spin degree of freedom is neglected from the model. (c)-(d) The mode amplitudes of the kink modes evaluated at $k_x=0.36$ with $\Delta_{\alpha}=0.2, t=1$; the amplitudes in different $\hat{s}_z$ components (up/down) are plotted separately.}
		\label{fig:Mapping}
	\end{figure}

	We found the low energy Hamiltonian to be 
	\begin{equation}
	H = \nu_{D}\left( q_{x}\hat{\tau}_{z}\hat{s}_{0}\hat{\sigma}_{x} + q_{y}\hat{\tau}_{0}\hat{s}_{0}\hat{\sigma}_{y} \right) + \Delta_D\hat{\tau}_{0} \hat{s}_{z}\hat{\sigma}_{0} +\Delta_{\alpha}\hat{\tau}_{0}\hat{s}_{x}\hat{\sigma}_{z}.
	\end{equation}
	We preform the unitary transformation $H\rightarrow \exp{(i\pi \hat{s}_y /4)} H \exp{(-i\pi \hat{s}_y /4)}$ which takes $\hat{s}_z\rightarrow \hat{s}_x$ and  $\hat{s}_x \rightarrow -\hat{s}_z$. After the transformation the Hamiltonian is now
	\begin{equation}
	H = \nu_{D}\left( q_{x}\hat{\tau}_{z}\hat{s}_{0}\hat{\sigma}_{x} + q_{y}\hat{\tau}_{0}\hat{s}_{0}\hat{\sigma}_{y} \right) + \Delta_D\hat{\tau}_{0} \hat{s}_{x}\hat{\sigma}_{0} -\Delta_{\alpha}\hat{\tau}_{0}\hat{s}_{z}\hat{\sigma}_{z}.
	\end{equation}
	Setting $\Delta_D = 0$ gives an identical Hamiltonian to the one presented by Li et al. \cite{li2013coupling}. Nonzero $\Delta_D$ in graphene can be interpreted as a Zeeman splitting caused by an external magnetic field along the $x-$axis, see Fig.~\ref{fig:Mapping}(a) for visualization of the mapping discussed in this section.
	
	\section{Mapping the tight binding model to AFM bilayer graphene}
	We consider the same unitary transformation as the preceding section. We then switch the order of the degrees of freedom in the Kronecker products of our Hamiltonian. Rewriting the Hamiltonian in this form we find 
	\begin{equation}
	H = \psi_{\mathbf{k}}^\dagger
	\begin{pmatrix}
	-\Delta_{\alpha} & \Delta_D & f_\mathbf{k} & 0 \\
	\Delta_D & \Delta_{\alpha} & 0 & f^*_\mathbf{k} \\
	f_\mathbf{k} & 0 & \Delta_{\alpha} & \Delta_D \\
	0 & f^*_\mathbf{k} & \Delta_D & -\Delta_\alpha 
	\end{pmatrix}
	\psi_{\mathbf{k}}.
	\end{equation}
	Writing $f_\mathbf{k} = \left| f_\mathbf{k} \right| \exp{(i\varphi_\mathbf{k})}$ we can transform the spinor vectors
	\begin{equation}
	\psi_{\mathbf{k}} =
	\begin{pmatrix}
	\psi_1 \\
	\psi_2 \\
	\psi_3 \\
	\psi_4 
	\end{pmatrix}
	\rightarrow
	\begin{pmatrix}
	\psi_1 \\
	\psi_2 \\
	e^{i\varphi_\mathbf{k}}\psi_3 \\
	e^{i\varphi_\mathbf{k}}\psi_4 \\
	\end{pmatrix},
	\end{equation}
	thus picking the phase factor of $f_\mathbf{k}$ from the Hamiltonian, leaving us with 
	\begin{equation}
	H = \psi_{\mathbf{k}}^\dagger
	\begin{pmatrix}
	-\Delta_{\alpha} & \Delta_D & \left| f_\mathbf{k}\right|  & 0 \\
	\Delta_D & \Delta_{\alpha} & 0 & \left|f_\mathbf{k}\right| \\
	\left|f_\mathbf{k}\right| & 0 & \Delta_{\alpha} & \Delta_D \\
	0 & \left|f_\mathbf{k}\right| & \Delta_D & -\Delta_\alpha 
	\end{pmatrix}
	\psi_{\mathbf{k}},
	\end{equation}
	which is the same Hamiltonian presented in Rakhmanov et al. \cite{rakhmanov2012instabilities} if we were to choose their $t_g$ parameter as 0, which is a reasonable approximation for AA-stacking BLG, and rotate $\mathbf{k}$ in $f_\mathbf{k}$ to fit their choice of reciprocal lattice unit cell orientation. We note that being able to write the Hamiltonian in this way proves why the conjugation of $f_\mathbf{k}$ has no effect on the edge spectra, as the phase plays no role in the form of the Hamiltonian. One can easily verify that transforming the Hamiltonian in the above form according to $H\rightarrow\hat{\sigma}_x\hat{s}_0 H \hat{\sigma}_x \hat{s}_0$ the sign of $\Delta_\alpha$ changes but the Hamiltonian is otherwise unchanged. Fig.~\ref{fig:Mapping}(b) depicts the mapping discussed in this section. In this physical representation of the Hamiltonian we can see that the bulk inversion operator $\mathcal{P}$ endows the edge states with modes that are symmetric and antisymmetric, as depicted in Fig \ref{fig:Mapping}(c-d) with regards to an inversion operation, which is the reason the kink modes do not anti-cross. For the case of Fig \ref{fig:Mapping}(c) each mode is perfectly localized in its own separate layer of the AA-BLG. For the case of Fig \ref{fig:Mapping}(d) the modes have non-zero amplitudes in the neighboring layer.
	
	\section{Symmetry of the kink modes in the first principles simulation}
\begin{figure}
		\centering
		\sidesubfloat[]{\includegraphics[width=0.3\columnwidth]{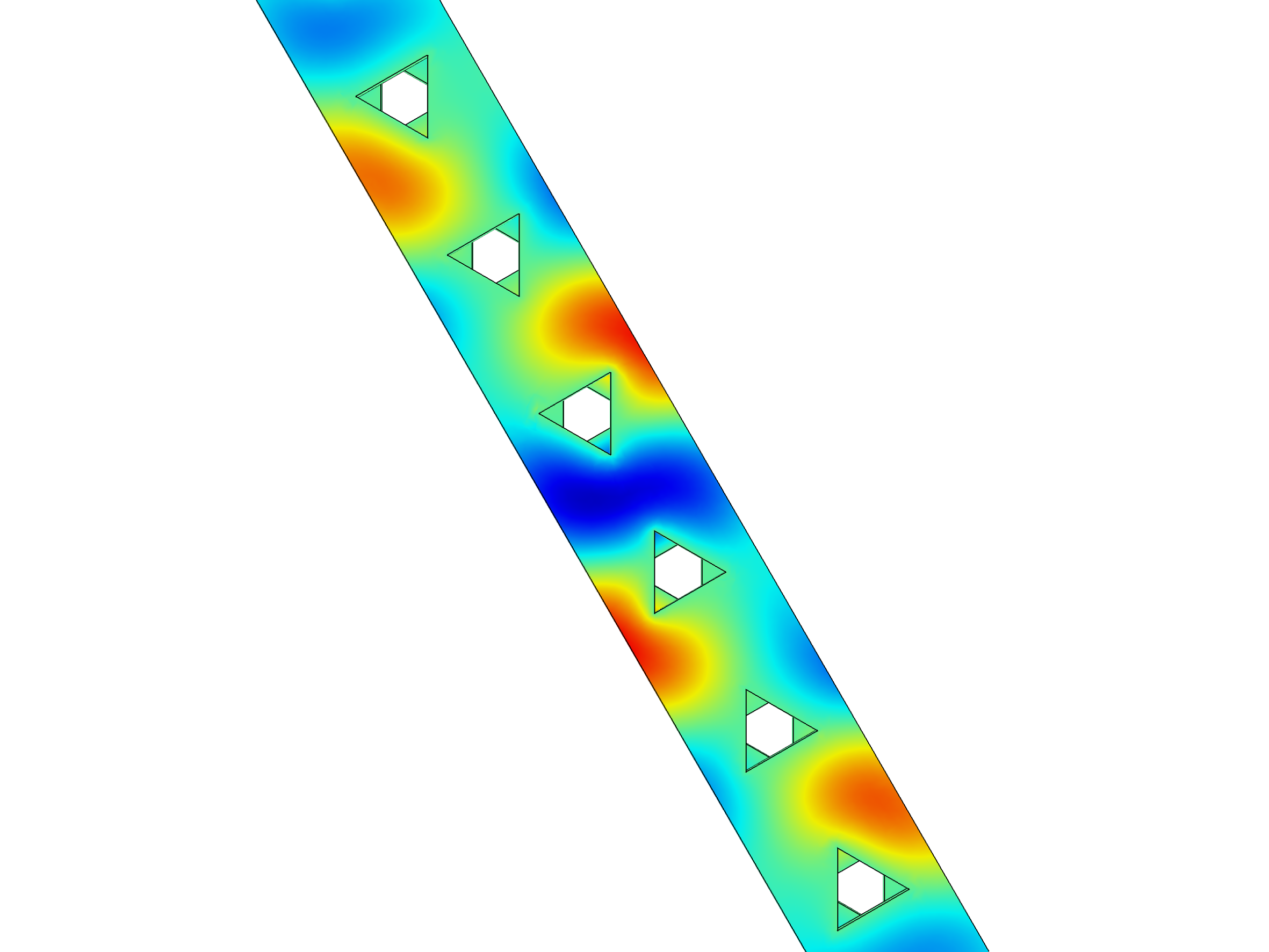}\label{fig:SupercellSymmetricModeNorm}}
		\sidesubfloat[]{\includegraphics[width=0.3\columnwidth]{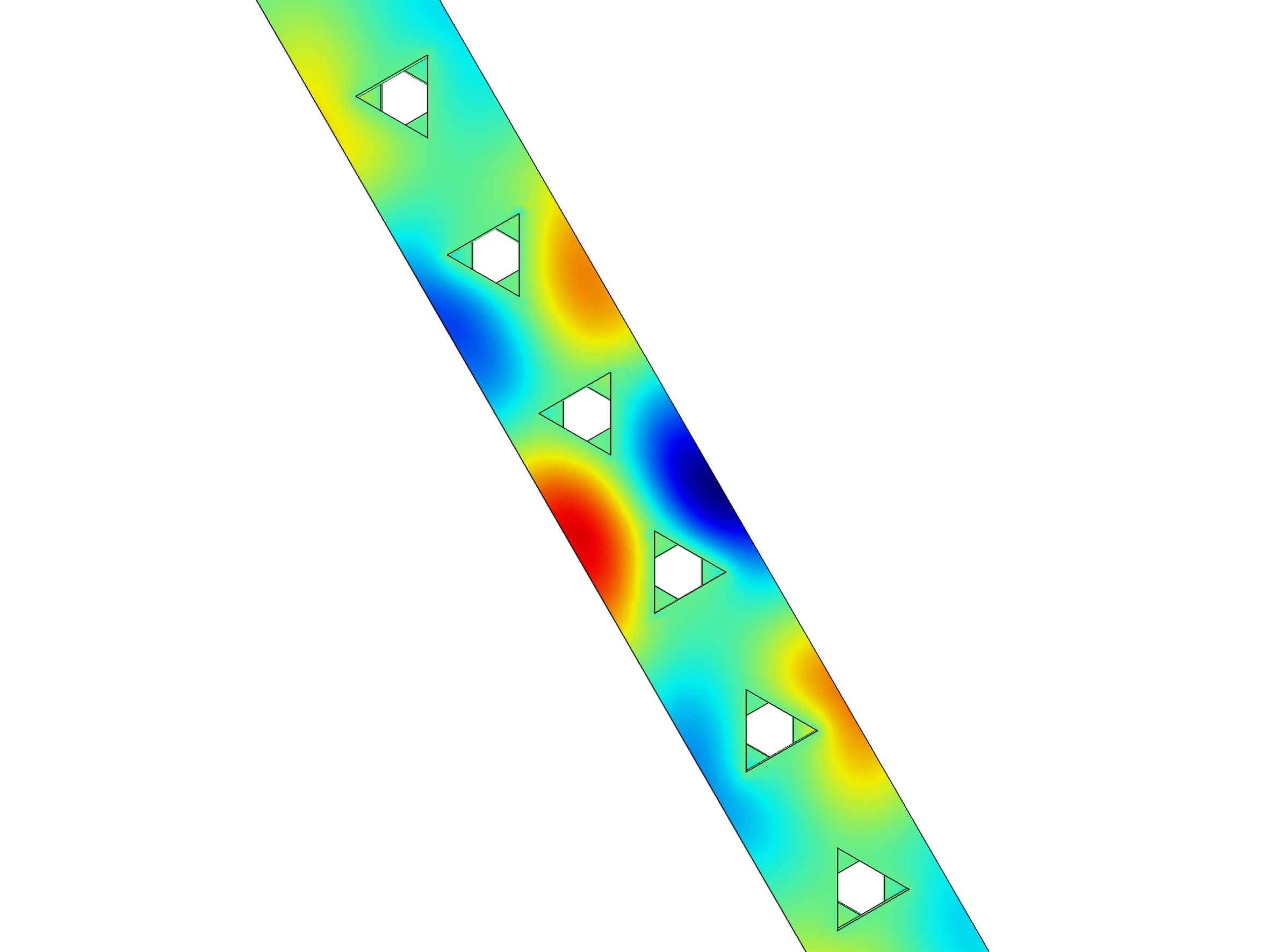}\label{fig:SupercellAntiSymmetricModeNorm}}

		\caption{\textbf{(a)} $|\mathbf{E}|$ profile at $z=3h/4$ for the symmetric supercell eigenmode for the SV-PTI with geometric parameters of Fig. 3(b) and $k_x = 2\pi/3a_0$, which is the projected $K'$ valley. \textbf{(b)} Same as (a) but for the antisymmetric mode.}
		\label{fig:SupercellModes}
	\end{figure}

	As discussed in the main text, the SV-PTI exhibits two kink modes at the interface between domains of different chirality. For the two kink modes to not couple to each other and open a mini band gap, they must possess opposite eigenvalues with regards to some symmetry operation. Due to the structure of the interface between the different SV-PTI domains, there is a spatial symmetry that is described by $(x,y,z)\rightarrow(x-a_0/2,-y,-z)$. This symmetry operation is known as a nonsymmorphic symmetry. While this symmetry is described by the full $\mathbf{E}$ vector field, it is difficult to visualize convincingly over a supercell. Thus, first, we verify that the kink modes calculated in the tight-binding model as described in the section VIII are indeed properly labeled as symmetric or antisymmetric modes. In Fig.~\ref{fig:Mapping}(c)-(d), it is clearly shown that the mode amplitudes show the symmetric(antisymmetric) behavior for the kink modes whose phase and group velocities have the same(opposite) sign(s), no matter whether $\Delta_D$ is vanishing or not. The only difference between the case with vanishing $\Delta_D$ (Fig.~\ref{fig:Mapping}(c)) and the case with nonzero $\Delta_D$ (Fig.~\ref{fig:Mapping}(d)) is whether $\hat{s}_z$ becomes a good quantum number; in Fig.~\ref{fig:Mapping}(c), two kink modes are not only labeled by symmetry/antisymmetry but also by $\hat{s}_z$ up/down, whereas $\hat{s}_z$ up and down components are mixed in the kink modes in Fig.~\ref{fig:Mapping}(d). Second, we plot the $E_z$ component from first-principle eigenmode simulation with the parameters from Fig. 3(d) at $k_x = 2\pi/3a_0$ and $z=3h/4$ in Fig. \ref{fig:SupercellModes}. We can observe that the $E_z$ component reveals that for th symmetric mode the field transforms back to itself while for the antisymmetric mode one must multiply by a factor of $-1$ for the field to transform back to its original value.
	
	\section{Finding $\Delta_D = 0$ in first principles simulation}
	To find the correct values of $R$, $\alpha$ and $h$ for which we get $\Delta_D = 0$ we note that in the tight binding model, for the case of $\Delta_D = 0$, the minimal separating frequency between bulk bands $\delta \omega$ is achieved exactly at the $K$ and $K^{\prime}$ points at the Brillouin zone (see Fig. 2 of the main text). We therefore optimized over $R$, $\alpha$ and $h$ for the position in momentum space of $\delta \omega$, which is found according to $d\omega / dk = 0$, such that it is achieved at the $K$ point. This assures us that $\Delta_D = 0$. This is verified by the linear dispersion of the edge modes (see Fig. 3(b)) and high broadband transmission (see Fig. 4(a)). In contrast, for $\Delta_D \neq 0$ the minimal separating frequency is achieved away from the $K/K^{\prime}$ points.

	\FloatBarrier
	
	\bibliographystyle{apsrev4-1}
	\bibliography{Main}
	
\end{document}